\documentclass[12pt]{article}
\usepackage{setspace}
\usepackage{amsmath,amssymb,mathrsfs}
\usepackage{color}
\usepackage{cite}
\usepackage{hyperref}
\usepackage{graphicx,epsfig}
\usepackage{appendix}
\usepackage{simpler-wick}
\usepackage{hyperref}

\usetikzlibrary{arrows.meta}

\begin{document}
\setlength{\topmargin}{-1cm} 
\setlength{\oddsidemargin}{-0.25cm}
\setlength{\evensidemargin}{0cm}
\newcommand{\e}{\epsilon}
\newcommand{\beq}{\begin{equation}}
\newcommand{\eeq}[1]{\label{#1}\end{equation}}
\newcommand{\bea}{\begin{eqnarray}}
\newcommand{\eea}[1]{\label{#1}\end{eqnarray}}
\renewcommand{\Im}{{\rm Im}\,}
\renewcommand{\Re}{{\rm Re}\,}
\newcommand{\diag}{{\rm diag} \, }
\newcommand{\Tr}{{\rm Tr}\,}
\def\draftnote#1{{\color{red} #1}}
\def\bldraft#1{{\color{blue} #1}}
\def\n{n \cdot v}
\def\ni{n\cdot v_I}
\begin{titlepage}
\begin{center}

\vskip 4 cm

{\Large \bf $T\bar   T$ Deformations through BRST Symmetry}

\vskip 1 cm

{E. de Sabbata$^{a}$~\footnote{E-mail:
\href{mailto:eliadesabbata@gmail.com}
{eliadesabbata@gmail.com}}, 
P.A. Grassi$^b$~\footnote{E-mail: \href{mailto:pietro.grassi@uniupo.it}{pietro.grassi@uniupo.it}},  M. Porrati$^c$~\footnote{E-mail: \href{mailto:mp9@nyu.edu}{mp9@nyu.edu}} }

\vskip .75 cm

{\em $^a$ Department of Mathematics, King's College London, 
Strand, London, WC2R 2LS, United Kingdom \\
$^b $ Dipartimento di Scienze e Innovazione Tecnologica (DiSIT),\\
Universit\`a del Piemonte Orientale, viale T. Michel, 11, 15121 Alessandria, Italy
\\
$^c$ Center for Cosmology and Particle Physics, \\ Department of Physics, New York University, \\ 726 Broadway, New York, NY 10003, USA}

\end{center}

\vskip 1.25 cm

\begin{abstract}
\noindent  
We study the $T\bar T$ deformation using its formulation as a CFT coupled to two-dimensional dynamical gravity.
Working within the BRST formalism, we apply the intertwiner construction of \href{https://arxiv.org/abs/2411.08865}{\texttt{arXiv:2411.08865}}  to obtain a unitary ``dressing'' map between undeformed CFT operators and elements of the BRST cohomology of the deformed theory.
We identify the resulting ``dressed'' operators corresponding to CFT primaries as the physical observables of the deformed theory and show that they arise from a field-dependent change of coordinates, in agreement with what is expected for the $T\bar T$ deformation.
We then give a non-perturbative definition of deformed correlation functions as BRST-invariant expectation values of dressed operators in the gauge theory.
Finally, we verify that our construction reproduces known structural and perturbative results.

\end{abstract}

\end{titlepage}
\tableofcontents
\newpage

\vskip 4 cm


\section{Introduction}
Generic irrelevant deformations of conformal field theories (CFTs) 
do not possess a well-defined high energy limit. A notable
exception is presented by the $T\bar T$ deformation~\cite{Smirnov:2016lqw,Cavaglia:2016oda}, which is exactly soluble even if its high-energy (UV)
completion is not a CFT fixed point, but instead shares several features with quantum gravity, such as a time delay growing with energy in 2 to 2 scattering, a minimal length scale, and a Hagedorn spectrum~\cite{Dubovsky:2012wk,Dubovsky:2017cnj,Camilo:2021gro}. In spite of this exotic UV behavior, the theory is integrable and can be obtained at the classical level via a field-dependent change of coordinates~\cite{Conti:2018tca}.  
The deformed theory also admits an infinite family of symmetries with associated conserved charges~\cite{Guica:2020uhm}, a structure that is argued to persist at the quantum level, at least perturbatively~\cite{Guica:2021pzy,Guica:2022gts}.  
In the large-$N$ case, the holographic dual description in $AdS_3$ with a finite bulk cutoff~\cite{McGough:2016lol} provides further evidence for the existence of these symmetries~\cite{Guica:2019nzm}.
In light of these properties, the $T\bar T$ deformation provides a controlled setting in which to probe the structure of irrelevant deformations, the emergence of gravitational behavior in quantum field theories (QFTs), and the dynamics of holography with finite cutoffs. \\

Even in the presence of these remarkable integrability properties, providing a fully non-perturbative UV-complete definition of $T\bar T$–deformed correlators and studying them beyond perturbation theory remains a difficult task that has recently received increasing attention. 
One of the main reasons behind this difficulty is the identification of the correct observables of the deformed theory.
These observables are inherently non-local, as suggested by the field-dependent change of coordinates that implements the deformation, or simply by the non-local nature of the deformed stress tensor.
An interesting approach was proposed in \cite{Cardy:2019qao}, where this non-locality is encoded by attaching infinitesimal strings to otherwise local operator insertions, and where point-splitting together with a non-local field renormalization is used to render the resulting correlators finite.
In a different direction, a useful perspective on the same problem is provided by the alternative description of the $T\bar T$
 deformation obtained by coupling the CFT to two-dimensional dynamical gravity~\cite{Callebaut:2019omt,Dubovsky:2018bmo}.
In particular, in \cite{Aharony:2023dod} a non-perturbative definition of deformed correlators is given in terms of a JT-gravity path integral; in this case, the resulting operators are defined at dynamical ``target-space'' positions, and their relation to the standard perturbative QFT operators is yet to be fully clarified. 
A formulation similar in spirit has also been given in terms of coupling the CFT to two-dimensional massive gravity \cite{Tolley:2019nmm}, providing a framework in which one can define finite-coupling correlators of “worldsheet’’ operators that can be directly matched with the perturbative QFT operators \cite{Hirano:2025tkq,Hirano:2025alr}; this approach generalizes earlier perturbative descriptions of the deformation in terms of a fluctuating random geometry \cite{Cardy:2018sdv,Hirano:2020nwq,Hirano:2020ppu}.
In particular, the massive gravity formulation allows for a very efficient computation of the all-order leading logarithmic corrections to the deformed correlators. Obtaining the subleading contributions, however, remains difficult due to the complications in determining the appropriate path-integral measure.
More generally, understanding correlators and other quantum features of $T\bar T$-deformed theories has become a very active line of research in recent years; for a partial list of additional relevant references, see \cite{He:2020udl,guica2020tt,He:2020qcs, He:2023kgq, Hirano:2024eab, Babaei-Aghbolagh:2024hti,Brizio:2024doe, He:2025ppz, Chen:2025jzb}.
\\

Given that the gauging of diffeomorphisms is at the heart of many non-perturbative constructions of $T\bar T$-deformed correlators, it is natural to investigate this subject using gauge-invariant methods, such as BRST quantization.
The main goal of this work is to employ such gauge-invariant techniques to describe the $T\bar T$-deformed observables and their deformed correlators.
In particular, we consider the formulation in which the deformation is implemented by coupling the CFT to two free massless scalar fields through the Virasoro constraints \cite{Callebaut:2019omt}.
In this picture, deformed observables are naturally described as elements of the BRST cohomology $H(\mathcal{Q})$, and correlators of such elements are automatically gauge-invariant.
The main difficulty lies in the fact that the BRST charge $\mathcal{Q}$ is a complicated object, and thus computing its cohomology is a challenging problem.
To deal with this issue, we employ a powerful technique recently developed in~\cite{Grassi:2024vkb}, which provides a general method for defining physical, gauge-invariant operators in gauge theories and gravity.
This allows us to gain practical control over the BRST cohomology $H(\mathcal{Q})$, to identify {a large class of} observables of the $T\bar T$-deformed theory within this cohomology, and to define their correlators.
Ultimately, the goal is to obtain a definition of these correlators that both matches perturbative QFT calculations and enables access to aspects of the non-perturbative regime. As we discuss below, the present work provides an important step in this direction.

\subsection*{Summary of the results}

In what follows, we provide a summary of our main results and outline the structure of the paper. \\

 Section~\ref{int1} briefly reviews the \emph{intertwiner} method introduced in~\cite{Grassi:2024vkb}, which we will later apply to determine the cohomology $H(\mathcal{Q})$.
This technique is particularly well adapted to our problem because it shows how to construct a {formally} unitary map $\Omega$, also called \emph{intertwiner}, that relates an ``undeformed'' BRST charge $\mathcal{Q}_0$ to the full ``deformed'' charge $\mathcal{Q}$ via $  \Omega\mathcal{Q} =  \mathcal{Q}_0 \Omega$.
The intertwiner operator $\Omega$ explicitly provides an isomorphism between the cohomologies $H(\mathcal{Q}_0)$ and $H(\mathcal{Q})$.
Usually, the undeformed charge $\mathcal{Q}_0$ is a simple object whose cohomology $H(\mathcal{Q}_0)$ is known, and the isomorphism can then be used to compute $H(\mathcal{Q})$ (for example, in our case a sector of the undeformed cohomology will just consist of undeformed CFT operators). 
The power of this method is that it does not require detailed knowledge of the full BRST charge; instead, it uses only a few simple properties of $\mathcal{Q}$  and $\mathcal{Q}_0$. \\

Section~\ref{ing} introduces our working example of the $T\bar T$ deformation, 
in which the undeformed theory is a generic CFT with central charge $c = 24$. 
This choice simplifies the analysis but is not restrictive. The extension to arbitrary central charge is in principle possible and will be discussed in a future work.
The deformation is implemented by coupling the CFT to two free scalar fields, interpreted as two-dimensional target-space coordinates, through the Virasoro constraints.
Section~\ref{ing} also defines the BRST charge $\mathcal{Q}$ associated with the gauge symmetry, as well as the corresponding ``undeformed'' charge $\mathcal{Q}_0$.
 \\

In Section~\ref{rop}, we introduce the additional ingredients required for the intertwiner method, such as the BRST anticharge operator \(\mathcal{R}\).
We then construct the intertwiner operator and present its explicit form in \eqref{RDA} and \eqref{intertwiners}.
\\

Section~\ref{vert} consists of a detailed analysis of the observables of the deformed theory, which belong to the cohomology of the exact
BRST charge $H(\mathcal{Q})$.
First of all, the cohomology of the undeformed charge $H(\mathcal{Q}_0)$ is explicitly computed in Subsection \ref{sec:q0cohom} and given in \eqref{holC}. 
In particular, the zero–ghost sector of $H(\mathcal{Q}_0)$ consists of the undeformed {local} CFT operators.
Therefore, the intertwiner $\Omega$ provides a unitary map between
the operator algebra of the CFT and a deformed operator algebra of elements in the cohomology $H(\mathcal{Q})$.
We will call this map a ``dressing,'' and we will identify the dressed CFT operators with {a large class of}  observables of the $T\bar T$-deformed theory.
Subsection~\ref{dresscft}, expresses the dressing in a closed form for the conformal primaries of the undeformed CFT.  
The dressed operators are explicitly given in \eqref{eq:primarytransfnew} and \eqref{eq:primarytransfgeneric}.
We find that the dressing map essentially coincides with the transformation that the operator undergoes under a change of coordinates given in \eqref{eq:trasfxpm2}, that depends on the auxiliary free scalars, in agreement with results obtained through various other formalisms.
In Subsection~\ref{vert}, we briefly discuss additional elements of the cohomology $H(\mathcal{Q})$ and explain how they can be obtained using the intertwiner.
\\

Deformed correlators for arbitrary values of the $T\bar T$ deformation parameter $\lambda$ are defined in Section~\ref{comp} as the BRST-invariant correlators \eqref{OBE} and \eqref{OBE2}.
These correlators can be easily shown to reduce to the undeformed CFT correlators in the $\lambda \to 0$ limit.
We then discuss how various results from the existing literature --such as flow equations and deformed conservation laws-- can be recovered within our framework.
Finally, as a {concrete} example, we expand the deformed two-point function for small $\lambda$ and verify that it matches known perturbative results.
\\

 Section~\ref{jjcohom} briefly discusses how to apply our construction to current--current $J\bar J$ deformations.  
The undeformed cohomology $H(\mathcal{Q}_0)$ and the deformed observables for $J\bar J$ deformations are {also} analyzed in Section~\ref{jjcohom}. \\ 

Finally, in Section~\ref{sec:conclusions}, we briefly discuss our results and their relevance, and indicate some possible future developments.
\\

Appendix~\ref{app:intzerocharge} discusses a simple generalization of the intertwiner construction introduced in Section~\ref{int1}, {which will be used in Appendix~\ref{app:B}}. \\
Appendix~\ref{app:B} contains a self-contained proof of the bosonic string no-ghost theorem based on the intertwiner operator.

\section{The intertwiner operator}\label{int1}
Here we briefly summarize the construction of an \emph{intertwiner} operator between two BRST charges, as presented in~\cite{Grassi:2024vkb}.
We start from a nilpotent BRST operator $\mathcal{Q}^B = \mathcal{Q}_0 + \mathcal{Q}_I$, where $\mathcal{Q}_0$ is itself nilpotent, $\mathcal{Q}_0^2 = 0$.
The nilpotency of $\mathcal{Q}^B$ then implies $\{ \mathcal{Q}_I, \mathcal{Q}_0 \} + \mathcal{Q}_I^2 = 0$.
Another essential ingredient is the \emph{anticharge} operator $\mathcal{R}$, satisfying
\begin{eqnarray}
    \label{opA1}
    \{\mathcal{Q}_0, \mathcal{R}\} = S\,, ~~~~~~ [S, \mathcal{Q}_0] =0 \,,
\end{eqnarray}
where  $S$ is a 
counting operator. 
We can construct an operator that intertwines between the ``simple'' BRST charge $\mathcal{Q}_0$ and the full charge $\mathcal{Q}^B$, provided that $\mathcal{Q}^B$ admits a finite decomposition into terms that, for an appropriate choice of the sign of $\mathcal{R}$, carry {positive} $S$-charge\,\footnote{See Appendix \ref{app:intzerocharge} for the generalization to the case that includes a vanishing $S$-charge term.}
\beq
\mathcal{Q}^B=\mathcal{Q}_0 +\!\!\sum_{0<n<N}\!\! \mathcal{Q}_n \, ,  \quad [S,\mathcal{Q}_n]=n\mathcal{Q}_n \, , \quad N\in \mathbb{N}\,.
\eeq{mass17}
The construction begins by introducing an auxiliary parameter $t$ and defining the $t$-dependent BRST charge
\beq
\mathcal{Q}(t)= \mathcal{Q}_0 + \mathcal{Q}_I(t)\,, \quad \mathcal{Q}_I(t)=\!\! \sum_{0<n<N} \!\! e^{-nt}  \mathcal{Q}_n\,, 
\eeq{mass18}
which interpolates from $\mathcal{Q}(0) = \mathcal{Q}^B$ at $t = 0$ to $\mathcal{Q}(+\infty) = \mathcal{Q}_0$ at $t = +\infty$. 
The nilpotency of the BRST charge, $(\mathcal{Q}^B)^2 = 0$, together with the commutation relation $[S, \mathcal{Q}_n] = n \mathcal{Q}_n$, 
implies that $\sum_{n+m=k} \{ \mathcal{Q}_n, \mathcal{Q}_m \} = 0$ for all $0 \leq k < 2N$, 
from which it follows that $\mathcal{Q}(t)$ is nilpotent for any $t$. 
Moreover, it also follows that $\{ \mathcal{Q}_0, \mathcal{Q}_I(t) \} + \mathcal{Q}_I(t)^2 = 0$. 
Using this last identity, we find
\bea
[\{\mathcal{Q}_I(t),\mathcal{R}\} ,\mathcal{Q}(t)] &=& \mathcal{Q}_I(t) \mathcal{R} (\mathcal{Q}_0+\mathcal{Q}_I(t)) + \mathcal{R} \mathcal{Q}_I(t) (\mathcal{Q}_0+\mathcal{Q}_I(t))  \nonumber \\ && -(\mathcal{Q}_0+\mathcal{Q}_I(t)) \mathcal{Q}_I(t)\mathcal{R}   -(\mathcal{Q}_0+\mathcal{Q}_I(t))\mathcal{R}\mathcal{Q}_I(t) \nonumber \\ &=&
\mathcal{Q}_I(t)\mathcal{R}\mathcal{Q}_0 +\mathcal{Q}_I(t)\mathcal{R}\mathcal{Q}_I(t) -\mathcal{R}\mathcal{Q}_0\mathcal{Q}_I(t) +\mathcal{Q}_I(t)\mathcal{Q}_0\mathcal{R}  \nonumber \\ && -\mathcal{Q}_0\mathcal{R}\mathcal{Q}_I(t) -\mathcal{Q}_I(t)\mathcal{R}\mathcal{Q}_I(t) \nonumber \\ &=&
[\mathcal{Q}_I(t), \{\mathcal{Q}_0,\mathcal{R}\}]= [\mathcal{Q}_I(t),S]=-\!\!\sum_{0<n<N} \!\!e^{-nt} n \mathcal{Q}_n=  {d\mathcal{Q}_I\over dt} \,.
\eea{mass14}
Next, we define an evolution operator $\Omega(t)$ by
\beq
{d\over dt} \Omega(t)=-  \Omega(t) \{\mathcal{Q}_I(t),\mathcal{R}\}\,.
\eeq{mass19}
Thanks to equation~\eqref{mass14}, we then find
\beq
{d\over dt}\Big( \Omega (t) \mathcal{Q}(t) \Omega^{-1}(t)\Big)= \Omega (t)\left( -[\{\mathcal{Q}_I(t),\mathcal{R}\} ,\mathcal{Q}(t)]  + {d\mathcal{Q}_I\over dt} \right) \Omega^{-1}(t)=0\,.
\eeq{mass20}
Integrating \eqref{mass20} from $t = 0$ to $t = +\infty$ with the boundary conditions 
$\Omega(t = +\infty) = 1$ and $\mathcal{Q}(t = +\infty) = \mathcal{Q}_0$, we find
\beq
\mathcal{Q}_0 -\Omega(0) \mathcal{Q}^B \Omega^{-1}(0) =0\,,
\eeq{mass21}
which indeed shows that $\Omega(0)$ is the intertwiner operator we wanted to construct.\\

For this construction, all the complexities hidden in $\mathcal{Q}_I$ are largely irrelevant, 
because the existence of an intertwiner depends only on $\mathcal{Q}_I$ having positive $S$-charge. 
Note that the construction of $\Omega$ requires the BRST charge to be nilpotent, and it does not work in the presence of gauge anomalies. \\

In the next sections, we will carry out this construction explicitly for the string description of the $T \bar{T}$ deformation first introduced in \cite{Callebaut:2019omt}. 
In practice, this will allow us to establish a direct map between undeformed and deformed operators, described in terms of the cohomology of $\mathcal{Q}_0$ and that of the full charge $\mathcal{Q}^B$, respectively.

\section{Worldsheet description of $T\bar T$ deformations}\label{ing}

In this section, we briefly review the worldsheet description of the $T \bar T$ deformation given in \cite{Callebaut:2019omt}, and introduce the ingredients necessary for our construction.
Our presentation differs from \cite{Callebaut:2019omt} only in that we take the worldsheet to be non-compact.
\\

We consider a two-dimensional CFT with stress-energy tensor $T_{\rm CFT}$ and central charge $c$. 
We then introduce two scalar fields $X^\pm$ with the free Lagrangian\,\footnote{In our convention, the metric is $ds^2=-d\tau^2+d\sigma^2$.} 
\begin{eqnarray}
    \label{caA}
    S =  
    S_{CFT} + \frac{1}{ 2 \pi \lambda} \int d\tau d\sigma \left(\dot X^+ \dot X^- - 
    X^{'+} X^{'-}\right),
\end{eqnarray}
where $\dot X^\pm \equiv \partial_\tau X^\pm(\tau,\sigma)$ and $ X'^\pm \equiv\partial_\sigma X^\pm(\tau,\sigma)$.
Assuming that the scalar fields are dimensionless, the deformation parameter $\lambda$ is also dimensionless.\footnote{Here we are referring to target space dimensions. With respect to these, all CFT operators and worldsheet coordinates are dimensionless.
}  
When this system is coupled to (classical) dynamical gravity, we obtain the classical Virasoro constraints
\begin{equation}
    \label{cab}
    \lambda \, T^{CFT}_{00} + \left(\dot X^+ \dot X^- + X^{'+} X^{'-} \right) = 0\,, \quad
   \lambda \, T^{CFT}_{01} + \left(\dot X^+  X^{'-} + X^{'+} \dot X^{-} \right) = 0\,,
\end{equation}
where $T^{\rm CFT}_{00}$ and $T^{\rm CFT}_{01}$ are the two independent components of the stress-energy tensor
of the CFT sector of the theory, defined as
\begin{equation}
T^{CFT}_{\mu \nu} \equiv \frac{4 \pi}{\sqrt{-g}} \frac{\delta S}{ \delta g^{ \mu \nu}}\,.
\end{equation}
The theory enjoys a target-space $ISO(1,1)$ invariance, consisting of shifts $X^\pm \to X^\pm + a^\pm$ and boosts $X^\pm \to e^{\pm \alpha} X^\pm$.
As we will later see in Section~\ref{comp}, this symmetry will turn out to correspond precisely to the Poincaré symmetry of the deformed theory. \\

To study this system, one can try to use the following gauge fixing for diffeomorphisms
\begin{eqnarray}
    \label{caB}
    \dot X^+ = X^{'+} + k^+\,, ~~~~~~~
    \dot X^ - = -X^{'-} + k^-\,,
\end{eqnarray}
Combining \eqref{caB} with the Virasoro constraints, we get
\begin{eqnarray}
    \label{caC}
    X^{'+} =  \frac{\lambda}{2 k^-} 
    \left( T^{CFT}_{00} + T^{CFT}_{01}\right) -\frac{k^+}{2}\,, ~~~~
     X^{'-} = -\frac{\lambda}{2 k^+} 
    \left( T^{CFT}_{00} - T^{CFT}_{01}\right)+\frac{k^-}{2}\,,
\end{eqnarray}
where $k^+$ and $k^-$ are constants,
and inserting back into the action \eqref{caA} we find 
\begin{eqnarray}
    \label{caD}
    S &=& S_{CFT} - \frac{\lambda}{4 \pi k^+ k^-}
    \int d\tau d\sigma (T^{CFT}_{00} - T^{CFT}_{01}) (T^{CFT}_{00} +T^{CFT}_{01}) + \frac{ k^+ k^-}{4 \pi \lambda} \int d\tau d\sigma
    \nonumber \\
    &=&
    S_{CFT} - \frac{\lambda}{4 \pi k^+ k^-}
    \int d\tau d\sigma \det(T^{CFT}_{\mu\nu}) - \frac{  k^+ k^-}{4 \pi \lambda} \int d\tau d\sigma\,.
\end{eqnarray}
The last term can be discarded, since it is just a constant.  
In \eqref{caD}, one can immediately recognize the $T \bar{T}$ deformation at first order.
Note, however, that the action \eqref{caA} is clearly not linear in $\lambda$, and therefore one should not expect the deformation to terminate at first order, {especially} at the quantum level. 
The apparent first-order form of \eqref{caD} arises only after performing a change of coordinates to implement the gauge-fixing condition \eqref{caB}. 
This transformation must explicitly depend on the initial configurations $X^\pm(\tau,\sigma)$, which are coupled to the CFT through the Virasoro constraint \eqref{caC}.
To make this point more explicit, one can add to $S_{CFT}$ a source term $\int d\tau d\sigma\, J(\tau,\sigma) \mathcal{O}(\tau,\sigma)$, where $\mathcal{O}$ is some CFT operator. 
In the quantum theory, one can then take functional derivatives with respect to $J$ to generate correlators. 
However, after performing the gauge-fixing transformation $(\tau,\sigma) \to (\tau',\sigma')$, one should consistently express the source in terms of the new coordinates {as a new function} $J(\tau',\sigma')$. 
This quantity now depends on the fields $X^\pm(\tau,\sigma)$, making it unclear how to proceed with the computation of correlators.
This is related to the fact that the $T\bar T$ deformation can be interpreted as a field-dependent change of coordinates \cite{Conti:2018tca, Guica:2022gts}.
In Section \ref{dresscft}, we will see that this idea also emerges naturally in our analysis.
\\

Since the gauge-fixing transformation is field-dependent and it is unclear how one should treat it at the quantum level, we will adopt a different approach.
The central charge of the CFT is $c$, and $c_X = 2$ for the $X^\pm$ system.  
If $c + c_X = 26$, the Weyl anomaly vanishes, allowing us to couple the system to gravity and quantize it \emph{before} choosing the gauge \eqref{caB}. 
Therefore, the simplest prototypical example is that of a CFT with central charge $c = 24$, and for the sake of clarity we will carry out our analysis in this case. 
The more general situation, in which the CFT has arbitrary central charge, is also possible and will be discussed in a future work.
As we show in the following sections, this quantization scheme enables precise definitions of deformed observables and correlators at arbitrary order in the deformation parameter $\lambda$. \\

First, we need to introduce the BRST charge
\begin{eqnarray}
    \label{BRSTA}
\begin{split}
       & {\mathcal Q} = i \int d\sigma \Big[ c^0 (  T^{CFT}_{00} + \frac{1}{\lambda} (\dot X^+ \dot X^- + X^{'+} X^{'-}) + \frac12 T_{00}^{gh}) \\
    & \hspace{2 cm}+ c^1 ( T^{CFT}_{01} + \frac{1}{\lambda}(\dot X^+ X^{'-} + X^{'+} \dot X^{-}) + \frac12 T_{01}^{gh})\Big]\,,
\end{split}
\end{eqnarray}
where the ghost fields $c^0$ and $c^1$ are associated, in the standard manner, with the two constraints
\begin{eqnarray}
    \label{BRSTB}
    \lambda \,T^{CFT}_{00} + \dot X^+ \dot X^- + X^{'+} X^{'-} + T_{00}^{gh} = 0 \,, \quad
    \lambda \,T^{CFT}_{01} +\dot X^+ X^{'-} + X^{'+} \dot X^{-}+ T_{01}^{gh} =0 \, ,
\end{eqnarray}
which are the Hamiltonian and momentum constraints.  
The additional terms in \eqref{BRSTA} represent the ghost contributions, and, for the moment, their explicit expressions are not relevant.
 \\
The theory is quantized using the following equal-time commutators
\begin{eqnarray}
    \label{etCA}
    &&[\dot X^+(\tau , \sigma), X^-(\tau, \sigma')] = - 2 \pi i  \lambda\, \delta(\sigma - \sigma')\,, ~~~~
    [\dot X^-(\tau, \sigma), X^+(\tau, \sigma')] =  -2 \pi i \lambda \, \delta(\sigma - \sigma')\,, ~~~~~~  \nonumber \\
    &&\{c^0(\tau, \sigma), b_0(\tau, \sigma')\} = - \delta(\sigma - \sigma')\,, ~~~~~~~~~~
    \{c^1(\tau, \sigma), b_1(\tau, \sigma')\} = - \delta(\sigma - \sigma') \, . 
\end{eqnarray}
The BRST variation of the $X^\pm$ fields is
\begin{eqnarray}
    \label{etCB}
    [{\mathcal Q}, X^\pm] = c^0 \dot X^\pm + c^1 X^{'\pm} = c^\mu \partial_\mu X^\pm .
\end{eqnarray}

Thanks to the equations of motion, we can always split the fields $X^\pm(\tau,\sigma)$ into their
 left- and right-moving components
\begin{equation}
    X^\pm(\tau,\sigma) = X_L^\pm(\tau-\sigma) + X_R^\pm(\tau+\sigma)\,.
\end{equation}
In terms of {these chiral fields}, the only non-zero commutation relations are
\begin{equation}\label{eq:appcommutators}
   \begin{split}
        &[\partial_\tau X^\pm_{L}(\tau-\sigma),X_L^\mp(\tau-\sigma')]=- [\partial_\sigma X^\pm_{L}(\tau-\sigma),X_L^\mp(\tau-\sigma')]=-\pi i\delta(\sigma-\sigma')\,, \\
         &[\partial_\tau X^\pm_{R}(\tau+\sigma),X_R^\mp(\tau+\sigma')]= [\partial_\sigma X^\pm_{R}(\tau+\sigma),X_R^\mp(\tau+\sigma')]=-\pi i \delta(\sigma-\sigma')\,.
   \end{split}
\end{equation}
In the following, we will often omit the subscripts, since it will be clear from the arguments whether a piece is left- or right-moving. \\

\begin{figure}[h]
\centering
\begin{tikzpicture}[scale=1.1, >=stealth]

  \draw[->] (-1,0) -- (4,0) node[right] {$\Re\!(z)$};
  \draw[->] (0,-1.6) -- (0,1.6) node[above] {$\Im\!(z)$};

  \draw[line width=1.2pt] (2,-1.4) -- (2,1.4);

  \filldraw (2,0.47) circle (2pt);
  \node[right] at (2.23,0.47) {$w_\tau$};

  \def\delta{0.10}  
  \def\r{0.15}      
  \def\h{0.47}      

  \draw[thin] (2+\delta,-1.4) -- (2+\delta,\h-\r);
  \draw[thin, domain=-90:90, samples=50] plot ({2+\delta+\r*cos(\x)}, {\h+\r*sin(\x)});
  \draw[thin] (2+\delta,\h+\r) -- (2+\delta,1.4);
  \draw[dashed, thin] (2+\delta,1.4) -- (2+\delta,1.8);
  \draw[dashed, thin] (2+\delta,-1.4) -- (2+\delta,-1.8);
  \draw[-{Triangle[angle=55:1.2mm]}] ({2+\delta+\r*cos(0)}, {0.05+\h+\r*sin(0)}) -- ++(0,0.001);

  \draw[thin] (2-\delta,-1.4) -- (2-\delta,\h-\r);
  \draw[thin, domain=90:270, samples=50] plot ({2-\delta+\r*cos(\x)}, {\h+\r*sin(\x)});
  \draw[thin] (2-\delta,\h+\r) -- (2-\delta,1.4);
  \draw[dashed, thin] (2-\delta,1.4) -- (2-\delta,1.8);
  \draw[dashed, thin] (2-\delta,-1.4) -- (2-\delta,-1.8);
  \draw[-{Triangle[angle=55:1.2mm]}] ({2-\delta+\r*cos(180)}, {-0.05+\h+\r*sin(180)}) -- ++(0,-0.001);

  \node[left] at (1.9,1.2) {\small $\gamma(\tau)$};

\end{tikzpicture}
\caption{ \small Contour $\gamma(\tau)$ around the slice $\Re\!(z)=\tau$; the black dot represents the pole at $z=w_\tau$.
}
\label{fig:contour}
\end{figure}
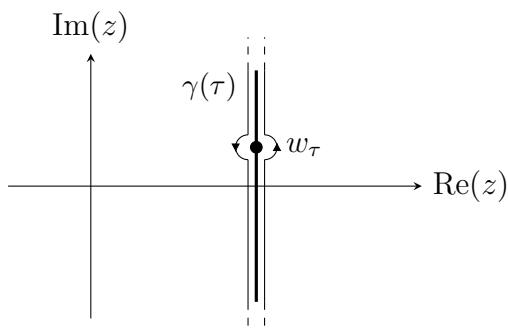

Instead of using equal-time commutators, we can rely on the OPE formalism.  
To this end, we first introduce the light-cone coordinates
\begin{equation}
    z=\tau-\sigma\,, \quad \bar z = \tau+\sigma\,.
\end{equation}
Next, we analytically continue $\sigma \rightarrow -i\sigma$ and introduce the OPEs for the holomorphic (left-moving) fields
\begin{eqnarray}
\label{eq:OPE1}
    \partial X^{\pm}(w)X^{\mp}(z) \sim \frac{\lambda}{w-z}\,, \quad c(w) b(z) \sim \frac{1}{w-z}\,,
\end{eqnarray}
with an analog formula holding for the antiholomorphic (right-moving) sector.
By treating $z$ and $\bar z$
 as complex coordinates, the usual OPE formalism can then be employed for practical computations, instead of relying on commutators.
In general, when all the charges decompose into purely holomorphic and antiholomorphic parts, one can compute them on an arbitrary closed contour and recover their action through Cauchy's formula.
However, in the following we will encounter operators defined as integrals over a spatial slice that do not decompose into purely holomorphic and antiholomorphic parts, yet we still wish to employ the OPE formalism. 
To illustrate how this can be done, let us consider the following example
\begin{equation}
    \begin{split}
        & C(\tau)  \equiv  \frac{i}{\pi} \int d\sigma \,X^-(\tau,\sigma) \partial_\tau X^+_L(\tau-\sigma)\,, \\
     & \big[C(\tau),f\big(X^-_L(\tau-\sigma)\big)\big] =  \lambda  X^-(\tau,\sigma) f' \big(X^-_L(\tau-\sigma)\big) \,,
    \end{split}
\end{equation}
which can be easily obtained using the commutation relation \eqref{etCA}.
Note that $C(\tau)$ cannot be split into $C(z)+\bar C(\bar z)$, as it contains terms that depend on both $z$ and $\bar z$.  
However, we still have the following result
\begin{equation}
\label{eq:generalisedcauchy1}
    \int_{\gamma(\tau)} \frac{dz}{2\pi i}\, X^-(z,\bar z) \partial_z X^+_L(z) f\big(X^-_L(w_\tau)\big)
    =  \lambda  X^-(w_\tau,\bar w_\tau) f'\big(X_L^-(w_\tau)\big)\,,
\end{equation}
where the integration contour $\gamma(\tau)$ surrounds the slice at fixed time $\tau$, with the point $w_\tau$ also lying on the same slice, as shown in Figure~\ref{fig:contour}. 
To show that \eqref{eq:generalisedcauchy1} holds, we use Stokes' theorem to generalize Cauchy's theorem to smooth functions
of $z$ and $ \bar z$
\begin{equation}\label{eq:cauchygeneralised2}
  \int_{\gamma(\tau)} \frac{dz}{2 \pi i} \frac{f(z,\bar z)}{z-w} \equiv \Big( \lim_{c \to \Re\!(w)^+} \!-\! \lim_{c \to \Re\!(w)^-}\!\Big)\!\int_{\Re\!(z)=c} \frac{dz}{2\pi i} \, \frac{f(z,\bar z) }{ z-w} =   f(w,\bar w)\,,
\end{equation}
where the limits correspond to the contributions from the segments of $\gamma(\tau)$ lying to the right and to the left of the slice, respectively, and the relation holds for sufficiently well-behaved functions $f$.\footnote{
More precisely, it holds whenever {$f(z,\bar{z})$ is smooth inside the domain of integration}. 
In particular, in our case all integrals involve only algebraic combinations of the fields $X^\pm(z,\bar z)$ and their derivatives, which can always be reduced to convergent sums of mode terms of the form $z^n \bar z^m$.
}
Indeed, the formula in \eqref{eq:cauchygeneralised2} can be proven by noting that on the slice $\Re\!(z)=c$, we have $f(z,\bar z)=f(z,2c-z)$.
If the analytic continuation of $f(z,2c-z)$ has no poles on the slice $\Re\!(z)=c$, then the difference of the integrals reduces to the residue at $z=w$, yielding the desired result.
\\

In particular, thanks to \eqref{eq:generalisedcauchy1}, we can still use the OPE to compute the action of non-holomorphic charge operators like $C(\tau)$ on local operators inserted at the same time $\tau$, using the {analog} of Cauchy's formula.\footnote{See also \cite{Hirano:2020ppu} for a similar approach.}  
We simply need to make the replacement
\begin{equation}\label{eq:integralsdz}
    	 \int d\sigma  \to \frac{1}{2} \int d \bar z \, -  \frac{1}{2} \int d z  \,,
\end{equation}
where the antiholomorphic part is included to account for contractions of the right-moving sector. \\

In holomorphic coordinates, up to an unimportant factor, the BRST charge is $\mathcal{Q} = Q + \bar Q$, with
\begin{eqnarray}
    \label{etCBA}
    Q = \int \frac{dz}{2 \pi i} \, c \,(T_{CFT} + \frac{1}{\lambda}\partial X^+ \partial X^-  - b \partial c) \,, ~~
    \bar Q = -\int \frac{d\bar z}{2 \pi i} \, \bar c \, ( \bar T_{CFT} + \frac{1}{\lambda}\bar\partial X^+ \bar\partial X^-  - \bar b \bar \partial \bar c) \,,
\end{eqnarray}
which are nilpotent independently (as the total charge vanishes) and mutually anticommute.
One important remark is that the charges in \eqref{etCBA} are always defined on constant time slices, as will be the case for all integrals we write below. We stress again that holomorphic coordinates are used for convenience, but all operators must lie on the same spatial slice in order to apply \eqref{eq:generalisedcauchy1} and similar formulas.
\\

To make contact with the classical constrained system discussed previously, we perform the shifts
\begin{eqnarray}
    \label{etCC}
    X^+ \rightarrow \hat X^+=X^+ +  \bar z\, \,, ~~~~~~
    X^- \rightarrow \hat X^- = X^- +  z\,,     
\end{eqnarray}
 Under this redefinition, the BRST charge acquires an additional term that we call $\mathcal{Q}_0$
\begin{eqnarray}
    \label{etCD}
  \mathcal{Q} \to \mathcal{Q}_0+\mathcal{Q}_1\,, \quad \quad {\mathcal Q}_0 
     = \frac{1}{\lambda}\bigg( 
        \int \frac{dz}{2 \pi i} \,
         c \, \partial X^ +  - \int \frac{d\bar z}{2 \pi i}\,
        \bar c \, \bar\partial X^-  \bigg)  \equiv  Q_0 + \bar Q_0 \,, 
\end{eqnarray}
where $\mathcal{Q}_1$ is the same as in \eqref{BRSTA}.  
The full {holomorphic} BRST operator is decomposed as follows: $Q = Q_0 + Q_1$, 
with $Q_0^2 = \{Q_0, Q_1\} = Q_1^2 = 0$ (and similarly for the anti-holomorphic sector).  
In addition, $\{Q, \bar Q\} = 0$.  
The subscripts $0$ and $1$ refer to the charge under the operator $S$, which will be introduced in the next section, following the general construction outlined in Section \ref{int1}.\\


We conclude this section with an observation.  
At the classical level, the deformation of the action is controlled by the following equation
\begin{eqnarray}
    \label{feA}
    \frac{d}{d \lambda} S =-    \frac{1}{2 \pi \lambda^2}\int d^2z \,\partial X^+ \bar \partial X^-\,,
\end{eqnarray}
The right-hand side of this equation is $\mathcal{Q}_0$-invariant. 
Indeed, we have
\begin{eqnarray}
    \label{holD}
    [Q_0, \partial X^+ \bar \partial X^-] =
    [\bar Q_0, \partial X^+ \bar \partial X^-]  = 0\,. 
\end{eqnarray}
 Moreover, it is also $Q_0$ and $\bar Q_0$ exact
\begin{eqnarray}
    \label{holE}
    \frac{1}{2 \pi \lambda^2} \int d^2z\, \partial X^+ \bar \partial X^- =  \frac{1}{2 \pi }\Big\{\bar Q_0, \Big[\ Q_0,\int d^2z \, b \bar b\Big] \Big\}\,.
\end{eqnarray}
Therefore, it is a trivial element of the $\mathcal{Q}_0$ cohomology.
We can then write \eqref{feA} as follow
\begin{eqnarray}
    \label{holEA}
     \frac{d}{d \lambda}S = - \frac{1}{2 \pi}\Big\{\bar Q_0, \Big[ Q_0,  \int d^2z \, b \bar b\Big] \Big\}\,.
\end{eqnarray}
In the Section \ref{rop}, we will explicitly construct an intertwiner operator $\Omega$ such that 
$\mathcal{Q}_0\, \Omega = \Omega\, \mathcal{Q}$.  
Together with the previous observation, this suggests that the deformation might be expressed as a redefinition of the operator basis.
Indeed, we can write
\begin{eqnarray}
    \frac{d}{d\lambda}S=-\Omega\, \Big\{\bar Q, \Big[ Q,  \Omega^{-1} \Big(\frac{1}{2 \pi}\int d^2z \, b \bar b \Big)  \Omega\Big] \Big\}\, \Omega^{-1}\,,
\end{eqnarray}
and thus the deformation is trivial after a ``change of basis’’ given by the intertwiner operator $\Omega$.
 \\

In the next section, we will extend this observation to the quantum level by introducing the anticharge operator $\mathcal{R}$ and analyzing the cohomology of the BRST operator $\mathcal Q$.

\subsection{$\mathcal{R}$ operator and intertwiner $\Omega$}\label{rop}


In this section, we explicitly construct the intertwiner operator $\Omega$ that relates the charges $\mathcal{Q}$ and $\mathcal{Q}_0$, given in \eqref{etCBA} and \eqref{etCD}, respectively. \\

The first step is to define the anticharge operator $\mathcal{R}$, which, for this theory, can be written as follows
\begin{eqnarray}
    \label{opAA}
    {\mathcal R} &=&\int d\sigma \, 
    \Big( -(b^0-b^1) \int d\sigma' \Theta(\sigma - \sigma') X^{'+}(\tau,\sigma') 
    +
    (b^0+b^1) \int d\sigma' \Theta(\sigma - \sigma') X^{'-}(\tau,\sigma')\Big)
    \nonumber \\
    &=& 
     \int d\sigma \, 
    \left(-(b^0-b^1)  X^{+}(\tau,\sigma)
    +
    (b^0+b^1) X^{-}(\tau,\sigma) \right)\,.
    ~~~~~~
  \end{eqnarray}
The two forms of the ${\mathcal R}$ operator given here are used to establish the relation with the construction of \cite{Grassi:2024vkb}.  
Using $\partial_\sigma \Theta(\sigma - \sigma') = \delta(\sigma - \sigma')$, we can remove the second integral in the first line of \eqref{opAA}.
Note that, unlike the BRST charge, the $\mathcal{R}$ operator defined in \eqref{opAA} depends on the time coordinate $\tau$. 
Moreover, the fields $X^{\pm}$ are ill-defined as two-dimensional local fields, and the same applies to any quantity in which they appear without $\partial$ or $\bar{\partial}$ derivatives. 
For these reasons, $\mathcal{R}$ is not a well-defined operator acting on the Hilbert space of the theory. 
Nevertheless, as we will show, it can still be used to define an outer automorphism of the observable algebra.
 \\

The operator $\mathcal{R}$ does not decompose into purely holomorphic and antiholomorphic parts. 
However, as explained in Section \ref{ing}, since we work at fixed time $\tau$, we can still employ the OPE formalism. 
It is convenient to split the anticharge into two pieces, $\mathcal{R} = R + \bar R$, as follows
\begin{eqnarray}
    \label{RA}
   R &=& -  \int \frac{dz}{2 \pi i} \, b(z) \, X^-(z,\bar z)\,, \nonumber 
  \\
   \bar R &=&   \int \frac{d\bar z}{2 \pi i} \,\bar b(\bar z) \, X^+(z,\bar z)\,.
\end{eqnarray}
We have chosen this way of splitting it so that $\{Q_0, \bar R\} = \{\bar Q_0, R\} = 0$.
The anticommutator between $Q_0$ and $R$, as well as its anti-holomorphic analogue, is given by
\begin{eqnarray}
    \label{RB}
\begin{split}
    & S \equiv \{Q_0, R\} = \int \frac{dz}{2 \pi i} \left(c(z)b( z)+ \frac{1}{\lambda} X^-(z,\bar z) \partial X^+(z)\right)\,, \\
 &\bar S \equiv \{\bar Q_0, \bar R\} =- \int \frac{d\bar z}{2 \pi i} \left( \bar c(\bar z) \bar b(\bar z)+ \frac{1}{\lambda} X^+ (z,\bar z)\bar \partial X^-(\bar z)\right)\,.
\end{split}
\end{eqnarray}
This implies $\{Q_0 + \bar Q_0, R + \bar R\} = S + \bar S$.
An important point is that, although $S$ and $\bar S$ contain the full fields $X^\pm(z,\bar z)$, one can equivalently replace them with their holomorphic part $X^-(z)$ and antiholomorphic part $X^+(\bar z)$, respectively.
Indeed, by going back to the $(\tau,\sigma)$ coordinates, one finds that the additional term arising from using the full $X^\pm$ fields in $S+\bar S$ is $\partial_\sigma\big(X^-_R(\tau+\sigma) X^+_L(\tau-\sigma)\big)$, which vanishes upon integration over a spatial slice. \\
The operator $S$ assigns $S$-charges of $+1$ to $c$ and $\partial X^{-}$, and $-1$ to $b$ and $\partial X^{+}$.  Similarly, $\bar S$ assigns $\bar S$-charges of $+1$ to $\bar{c}$ and $\bar{\partial}X^{+}$, and $-1$ to $\bar{b}$ and $\bar{\partial}X^{-}$.
Using these charges, we can decompose the BRST operators as $Q = Q_{0} + Q_{1}$ and $\bar{Q} = \bar{Q}_{0} + \bar{Q}_{1}$, thus satisfying the assumptions required for the construction of the intertwiner described in Section~\ref{int1}.
\\

Since the $\mathcal{R}$ operator correctly generates a counting operator $S+\bar S$ with the desired properties, we can use it to compute the intertwiners $\Omega(t)$ and $\bar{\Omega}(t)$ by solving the following differential equation \eqref{mass19}, which in our case is 
\begin{eqnarray}
    \label{RC}
    \Omega^{-1} \frac{d}{dt} \Omega = - e^{-t}\{Q_1+\bar Q_1, R+\bar R\}\,.
\end{eqnarray}
Again, one can check that $\{Q_1, \bar R\} = \{\bar Q_1, R\} = 0$.  
Therefore, it is convenient to define
\begin{equation}
    \label{RD}
 \begin{split}
        &\Sigma \equiv \{Q_1, R\} = \int \frac{dz}{2 \pi i}\,  \left( 
    :T X^-: +  \, :c b: \, \partial X^- \right) \,, ~~~~~
   \\
    &\bar \Sigma \equiv \{\bar Q_1, \bar R\} =-  \int \frac{d\bar z}{2 \pi i} \, \left( 
    :\bar T X^+: + \,   :\bar c \bar b: \,\bar \partial X^+ \right) \,, ~~~~~
 \end{split}
\end{equation}
where, again, the underived $X^\pm$ operators include both the holomorphic and antiholomorphic parts.
The normal ordering is needed since $T$ contains the $\partial X^+ \partial X^-$ term.
In equation \eqref{RD}, the complete $T$ tensor is given by 
\begin{equation}
    \begin{split}
         \label{RDA}
    T = T_{CFT} +  \frac{1}{\lambda}:\partial X^+ \partial X^-: - 2 :b \partial c: - :\partial b c:\,,  \\
    \bar T = \bar T_{CFT} + \frac{1}{\lambda}:\bar \partial X^+ \bar \partial X^-: - 2 :\bar b \bar\partial \bar c: - :\bar\partial \bar b \bar c:\,.
    \end{split}
\end{equation}
As was the case for $R$ and $\bar R$, the operators $\Sigma$ and $\bar{\Sigma}$ depend on the choice of slice and are not well-defined operators on the Hilbert space. \\
Note that the second term in each $\Sigma$ is related to the ghost currents $J = c b$ and $\bar{J} = \bar{c} \bar{b}$. 
By performing an integration by parts, we can also rewrite these expressions as 
\begin{equation}\label{RE}
\begin{split}
    &\Sigma = \int 
    \frac{d z}{2 \pi i} (T - \partial J) \bigg(\int  \frac{dy}{2 \pi i}\,G(|z-y|)\, 
  \partial X^-(y)+\int  \frac{d \bar y}{2 \pi i}\,G(|z-y|)\, 
  \bar \partial X^-(\bar y)
  \bigg)
    \,, ~~~~~ \\
    &\bar \Sigma =
    \int \frac{d\bar z}{2 \pi i} 
    (\bar T - \bar \partial \bar J) \bigg(\int  \frac{dy}{2 \pi i}\,G(|z-y|)\, 
  \partial X^+(y)+\int  \frac{d \bar y}{2 \pi i}\,G(|z-y|)\, 
  \bar \partial X^+(\bar y)
  \bigg)  \,,
\end{split}
\end{equation}
where the Green's function is $G(|z - y|) = -\log(\kappa |z - y|)$, and $\kappa$ is a free, undetermined parameter. \\

Twisted versions of $T$ and $\bar{T}$ naturally emerge; namely
\begin{equation}
    \label{RDAA}
   \begin{split}
        T - \partial J = T_{CFT} \,+   \frac{1}{\lambda}:\partial X^+ \partial X^-: -  :b \partial c:\,, \\
    \bar T - \bar \partial \bar J= \bar T_{CFT} \, +  \frac{1}{\lambda}:\bar \partial X^+ \bar \partial X^-: - :\bar b \bar\partial \bar c: \,.
   \end{split}
\end{equation}
This implies that the sector $X^\pm, c,b, \bar c, \bar b$ is a pure topological model. Indeed, the central charge of $T - \partial J,  \bar T - \bar \partial \bar J$ is $(c_{CFT}, \bar c_{CFT})$ since the central charges of $X^\pm$ cancel the central charges of the ghost system. The conformal dimension of $b,c$ and their antiholomorphic partner is now $-2$ since they form a $(1,0)$ $bc$ system instead of a $(2,-1)$ ghost system. This means that the system $X^\pm, c,b, \bar c, \bar b$ is an auxiliary system introduced to describe the deformations, but {it contains} no physical degrees of freedom.\footnote{This is very similar to the discussion in \cite{Dubovsky:2018bmo} using topological gravity and in \cite{Dubovsky:2023lza} in the case of $J\bar J$ deformations.} 
  \\

Explicitly, the intertwiner operator that solve equation \eqref{RC} is
\begin{eqnarray}\label{intertwiners}
    \Omega(t)= \exp \!\big( e^{-t} (\Sigma + \bar \Sigma)\big)\,, \quad \Omega \equiv \Omega(0)=  \exp \!\big(\Sigma + \bar \Sigma\big)\,.
\end{eqnarray}

In the next sections, we will first compute the cohomology $H(\mathcal{Q}_0)$ of the charge $\mathcal{Q}_0$, and then use the intertwiners \eqref{intertwiners} to relate it to the cohomology of the full BRST charge, $H(\mathcal{Q})$.  
Importantly, observables in the cohomology of the full {BRST} charge form correlators that match those of the $T \bar T$-deformed theory.\\

\section{Observables in the deformed theory}\label{vert}

\subsection{The cohomology of $\mathcal{Q}_0$}\label{sec:q0cohom}

To compute the operator algebra of the theory, we need to determine the BRST cohomology of $\mathcal{Q}$.  
The method introduced in \cite{Grassi:2024vkb} and outlined in Section~\ref{int1} first computes the cohomology of $\mathcal{Q}_0$, namely $H(\mathcal{Q}_0)$, and then uses the intertwiner $\Omega$ to obtain the full cohomology. \\
For each class $[U] \in H(\mathcal{Q}_0)$, we choose a representative $U$, and the corresponding \emph{dressed} operator $\Omega^{-1} U \Omega$ represents a class in $H(\mathcal{Q})$. 
In other words, the intertwiner operator $\Omega$ explicitly provides an isomorphism between the two cohomologies
\begin{equation}\label{eq:isocohom}
    [U] \in H(\mathcal{Q}_0) \  \Leftrightarrow \ [\Omega^{-1} U\Omega] \in H(\mathcal{Q}) \,.
\end{equation}

We begin by determining the cohomology $H(\mathcal{Q}_0)$. \\
Given the BRST operator $\mathcal Q_0 = Q_0 + \bar Q_0$ as 
\begin{eqnarray}
    \label{holA}
    {\mathcal Q}_0 = \frac{1}{\lambda} \int \frac{dz}{2 \pi i} \,  c \,\partial X^+ -
\frac{1}{\lambda}     \int \frac{d\bar z}{2 \pi i}  \, \bar c \, \bar\partial X^-\,,
\end{eqnarray}
we can compute its action on the on-shell fields 
\begin{eqnarray}
    \label{holB}
    [Q_0,  X^-] =  c\,, ~~~~~\{Q_0, c\} = 0\,, ~~~~~
   \{{Q}_0, b\} = \frac{1}{\lambda} \partial X^+\,, ~~~~~
     [Q_0, X^+] =0\, .
\end{eqnarray}  
This type of BRST symmetry is a topological symmetry (see, for example, \cite{Witten:1988xj, Witten:1991zz, Hori:2003ic}, where it appears as the topological B-model).  
In the previous section, we observed that in the intertwiner only a particular combination of the stress-energy tensor $T$ and the ghost current $J$ enters, indicating that even from the conformal field theory point of view the fields $X^\pm$ together with the ghost systems form a topological, non-physical sector. \\

The BRST cohomology of $\mathcal{Q}_0$ is given by representatives $U$ that satisfy
\begin{eqnarray}
 \label{newpAB}
   [Q_0, U] = 0\,, ~~~~~ [\bar Q_0, U] =0\,,
\end{eqnarray}
The latter are defined up to $\mathcal{Q}_0$-exact pieces, 
namely 
\begin{equation}
   U \sim U + \delta U\,, \quad  \delta U = \{\mathcal{Q}_0, \Lambda\}\,,
\end{equation}
where $\Lambda$ is an arbitrary composite of local fields.
The general solution of \eqref{newpAB} is generated by the vertex operators $(1, c \, \delta(X^-))$ in the holomorphic sector and by $(1, \bar c \, \delta(X^+))$ in the anti-holomorphic sector, yielding the following
cohomology 
\begin{eqnarray}
    \label{holC}
    H(\mathcal{Q}_0) = \Big\{\mathcal{A}\,,\,   \mathcal{B}\, c\, \delta(X^-)\,, \, 
    \mathcal{C}\, \bar c \,\delta(X^+)\,, \, 
    \mathcal{D}\, c \,\bar c :\!\delta(X^-) \delta(X^+)\!: \!\Big\} \,.
\end{eqnarray}
The coefficients 
$\mathcal{A}(z,\bar z), \, \mathcal{B}(z,\bar z), \, \mathcal{C}(z,\bar z), \, \mathcal{D}(z,\bar z)$ are local fields of the CFT degrees of freedom.  
In the last term, normal ordering has been used to account for the fact that the fields $X^\pm$ have a non-trivial OPE with each other.
Note that the terms containing Dirac delta functions, such as $\mathcal{B} \, c \, \delta(X^-)$, can be formally written as $\mathcal{B} \, c \, \delta(X^-) = [Q_0, \mathcal{B} \, \Theta(X^-)]$. 
However, the Heaviside distribution $\Theta(X^-)$ lies outside our Hilbert space since it does not have compact support, and therefore these terms cannot be regarded as exact (see also~\cite{Friedan:1985ge}). \\
An interesting point is that in the Hilbert space there should always be pairs of cohomology classes related by conjugation:
$|\omega_0\rangle = |0\rangle$ and $|\omega_1\rangle = c\,\delta(X^-)\,|0\rangle$, 
so that $
\langle\omega_0| \omega_1\rangle 
= \int dX^- \, dc \,\langle 0|\, c\,\delta(X^-)\,|0\rangle = 1 \, $. 
Note that the field $X^-$ has positive $S$-charge $+1$, whereas $\delta(X^-)$ has negative $S$-charge $-1$. 
The entire cohomology \eqref{holC} is therefore restricted to operators with vanishing $S$- and $\bar S$-charges. 
This also follows from the algebraic Poincar\'e lemma: for non-vanishing $S$ or $\bar S$ charge, any closed operator is also exact. 
\\

For a general worldsheet theory, not every element of the cohomology corresponds to a physical operator, and thus a priori it is not guaranteed that all dressed versions of the operators in the $\mathcal{Q}_0$ cohomology give rise to physical vertex operators.
For instance, in worldsheet theories one typically encounters unintegrated vertex operators of the form
\begin{equation}
    U = c\, \bar c\, V \,,
\end{equation}
where $V$ is a ghost-free primary of weights $(h,\bar h) = (1,1)$.
This type of operator can be obtained by Fourier transforming the elements in the ghost $(1,1)$ sector of \eqref{holC},\,\footnote{We refer to ghost number $(a,b)$ with respect to the left- and right-moving ghost number currents $J = :\!bc\!:$ and $\bar J = :\!\bar b \bar c\!:$.}  and choosing the momenta so as to obtain the desired conformal weights.
These operators are usually regarded as the physical ones. \\
If $U$ is an element of the cohomology, then it can be converted into a ghost-free integrated vertex operator $V$, which is guaranteed to be an element of the cohomology of $\mathcal{Q}_0 / (\partial, \bar\partial)$; that is, it is BRST closed and not exact up to total derivatives.
Given an integrated vertex operator $V$, the insertion $\int d^2 z\, V$ is also regarded as a physical insertion. \\
The integrated and un-integrated operators are related by the descent equations
\begin{eqnarray}
    \label{newpA}
    [Q_0, V] = \partial W\,, ~~~~~
    [\bar Q_0, V] = \bar\partial\, \overline{W}\,, ~~~~~
    \{Q_0, \overline W \} = \partial U\,, ~~~~~
    \{\bar Q_0, W\} = \bar\partial U\,.
\end{eqnarray}
We can also convert un-integrated vertex operators into integrated ones by using the fields $B_0$ and $\bar B_0$, defined as
\begin{eqnarray}
    \label{newpC}
    B_0 =  b\, \partial X^-\,, ~~~~~
    \bar B_0 = \bar b\, \bar\partial X^-\,.
\end{eqnarray}
The operators $B_0$ and $\bar B_0$ have the following BRST transformations:
\begin{eqnarray}
    \label{newpD}
    \{Q_0, B_0\} =  \frac{1}{\lambda}:\!\partial X^+\partial X^-\!: - :\!b\partial c\!:\,, \qquad
    \{\bar Q_0, \bar B_0\} =  \frac{1}{\lambda}:\!\bar\partial X^+\bar\partial X^-\!: - :\!\bar b \bar\partial \bar c\!:\,.
\end{eqnarray}
The terms on the right-hand side of these equations are the twisted energy-momentum tensors of the \emph{undeformed} system $X^\pm, b, c, \bar b, \bar c$. \\


In the cohomology \eqref{holC}, there is an additional family of solutions: these are the CFT operators $\mathcal{A}(z,\bar z)$.
These operators have zero ghost number and belong to the actual BRST cohomology (not only up to total derivatives).
These solutions are somewhat unusual, as they do not appear, for example, in the cohomology of bosonic string theory.
Moreover, their conformal weights depend on the details of the undeformed CFT, and once they are dressed using the intertwiner operator, they will no longer possess definite weights (rather, as we will see, they become combinations of operators with different weights).
Nevertheless, although these operators may at first seem unphysical, they are the natural candidates for observables in the $T\bar T$-deformed theory, since they are in one-to-one correspondence with the observables of the undeformed CFT.
For this reason, in Section \ref{dresscft} we will focus on the problem of finding a closed formula for these dressed observables, and in Section \ref{comp} we will \emph{define} the $T \bar T$-deformed correlators in terms of them.
\\


We conclude this section with a remark.
The introduction of the $X^\pm$ and $b,c$ systems essentially reproduce the $b,c$ and $\beta,\gamma$ system for two-dimensional twisted topological gravity, therefore we have to take into account that the Hilbert space spanned by $X^\pm$ might have several equivalent copies (at different pictures). 
For example, instead of $ U = c \bar c  :\!\delta(X^-) \delta(X^+)\!:$ one can consider the vertex operator 
\begin{eqnarray}
    \label{PIA}
     U  &= \partial c \bar c  :\!\delta(\partial X^-) \delta(X^+)\!:\,,
\end{eqnarray}
which is still in the cohomology of 
$\mathcal{Q}_0$ since $[Q_0, \partial X^-] =  \partial c$. Or for example
\begin{eqnarray}
    \label{PIB}
 U  &= c\partial c \bar c  :\!\delta(X^-)\delta(\partial X^-) \delta(X^+)\!:\,,
\end{eqnarray}
 which has a higher ghost number. In order to change the picture, there are suitable operators known as Picture Changing Operators, such as 
 \begin{eqnarray}
     \label{PIC}
     Z = b \delta(\partial X^+) \bar b \delta(\bar \partial X^-) \,,
 \end{eqnarray}
{This operator maps $U$ into} the first element of the cohomology \eqref{holC}
\begin{eqnarray}
    \label{PID}
    Z \left( \mathcal{D} c\delta(X^-) \bar c \delta(X^+) \right) \rightarrow \mathcal{D}\,,
\end{eqnarray}
 where $\mathcal{D}$ is a field of the CFT. 
 
 \subsection{Dressing of CFT primaries}\label{dresscft}
 
In this section, we address the problem of finding a closed formula for dressed fields in the special case where they are conformal primaries of the “matter’’ CFT.
The matter CFT primaries correspond to the ghost–zero elements of the cohomology $H(\mathcal{Q}_0)$ given in \eqref{holC}. 
The dressed field is obtained by using the intertwiner given in \eqref{intertwiners}
\begin{eqnarray}
    \label{DDFA}
    {\mathcal O}^\Omega(z,\bar z) \equiv \ :\!\exp\!\big(\!-\!({ \Sigma+ \bar \Sigma)\big)} \, {\mathcal O}(z,\bar z) \,\exp\!{\big((\Sigma + \bar \Sigma)\big)}\!:\,.
\end{eqnarray}
Note that the operator $\mathcal{O}(z,\bar z)$ is always inserted at time $\tau$, and that the intertwiner $\Omega= \exp (\Sigma+\bar \Sigma)$ is likewise defined as an integral over the constant $\tau$ slice. 
As stated in \eqref{eq:isocohom}, the dressed fields defined in \eqref{DDFA} are the ghost–zero elements of the cohomology of the full BRST charge $H(\mathcal{Q})$, since they are obtained by conjugation with the intertwiner $\Omega$. 
These dressed fields will be identified with the observables of the $T\bar T$–deformed theory.
\\

Formula \eqref{DDFA} can be expanded in terms of the nested commutators $[\Sigma + \bar \Sigma, \mathcal{O}]_n$ as follows
\begin{eqnarray} \label{DDFA2}
    {\mathcal O}^\Omega(z,\bar z)  = 
    \sum_{n=0}^\infty \frac{(-1)^n}{n!} \,[\Sigma +\bar \Sigma, \mathcal{O}]_n\,.
\end{eqnarray}
In this section, to simplify the notation, we will often omit the arguments of the operators, which are always understood to be $z$ and $\bar{z}$. Moreover, we will often omit the normal-ordering symbol; normal ordering is always implicitly assumed for products of non-commuting operators evaluated at coincident points.
We begin by considering the first commutator for an operator $\mathcal{O}(z, \bar{z})$ with conformal weights $h$ and $\bar{h}$\,
\begin{eqnarray}\label{eq:firstcommutator}
    [\Sigma +\bar \Sigma, \mathcal{O}]= \Big(h \,\partial X^- \mathcal{O} + X^-\partial \mathcal{O} + \bar h \,\bar \partial X^+\mathcal{O}+X^+\bar \partial \mathcal{O}\Big)\,.
\end{eqnarray}
This formula already suggests that the commutator $[\Sigma(\bar{z}) + \bar{\Sigma}(z), \cdot]$ acts as the generator of a coordinate transformation that depends on $X^-(z,\bar z)$ and $X^+(z,\bar z)$. As we will see later, this is indeed the case: the finite transformation is completely fixed by the infinitesimal one, as usual.
To compute higher-order commutators, we first need the following relations
\begin{equation}\label{eq:basiccommutators}
\begin{split}
        &[\Sigma+\bar \Sigma,X^-]=\big(X^-\partial X^-+X^+\bar \partial X^-\big)\,, \\
    &[\Sigma+\bar \Sigma,X^+]=\big(X^-\partial X^++X^+\bar \partial X^+\big)\,.
\end{split}
\end{equation}
Crucially, this transformation mixes the holomorphic and antiholomorphic sectors in a non-trivial way. 
Indeed, even if we start with a holomorphic operator $\mathcal{O}(z)$ with $\bar{h} = 0$, we see from \eqref{eq:firstcommutator} that there is a term $X^- \partial \mathcal{O}$, which depends on the antiholomorphic coordinate $\bar{z}$. 
Moreover, due to the first commutator in \eqref{eq:basiccommutators}, higher-order commutators $[\Sigma + \bar{\Sigma}, \mathcal{O}(z)]_n$ will contain $X^+$. 

\paragraph{Warm-up example.} To understand the structure of these nested commutators, we first consider a somewhat simplified version of the problem. 
Let us set $X^+ = 0$ and take a holomorphic operator with conformal dimension $h = 0$.
In this way, the holomorphic sector does not mix with the antiholomorphic one. 
Although the result will not be complete, it will help us formulate an \emph{ansatz} that we will later verify.
The commutators in \eqref{eq:basiccommutators} reduce to
\begin{eqnarray}
    \label{DDFB}
    [\Sigma, \partial X^-] =  \partial(X^- \partial X^-)\,, \quad
    [\Sigma, X^-] =  X^- \partial X^- \,.
\end{eqnarray}
Moreover, one can check that
\begin{eqnarray}
    \label{DDFC}
    [\Sigma, \partial X^-]_n =  \partial^n \left(\left(X^-\right)^n \partial X^- \right)\,.
\end{eqnarray}
For a holomorphic field $\mathcal{O}$ of conformal weight $h$, we have
\begin{eqnarray}
    \label{DDFE}
   && [\Sigma, {\mathcal O}] = (h -1) \partial \!\left( X^- \right) {\mathcal O} + 
     \partial\! \left( X^- {\mathcal O}\right) \,,\\
      && [\Sigma, [\Sigma, {\mathcal O}]]
      = 
    \Big( h (h -1)(\partial X^-)^2 \mathcal{O}+2(h-1)\partial \big(X^-\partial X^- \mathcal{O}\big)+\partial^2 \big((X^-)^2\mathcal{O}\big) \Big) \nonumber\,.
\end{eqnarray}
Proceeding further, we find that
\begin{eqnarray}
    \label{DDFG}
    [\Sigma, {\mathcal{O}}]_{n} = (h -1) F[X^-, {\mathcal{O}}] 
    + \partial^n \left(\left(X^- \right)^n {\mathcal{O}} \right)\,,
\end{eqnarray}
Therefore, for a conformal weight $h = 1$, we obtain a simple recursive formula
\begin{eqnarray}
    \label{DDFH}
    [\Sigma, {\mathcal{O}}]_{n} = \partial^n \left( \left(  X^- \right)^n {\mathcal{O}} \right)\,.
\end{eqnarray}
Using this formula, the dressed operator $\mathcal{O}^\Omega$ is given by
\begin{eqnarray}
\label{DDFI}
\mathcal{O}^\Omega = \sum_{n=0}^\infty \frac{(-1)^n}{n!} \,
     \partial^n \!\left(\!\left(X^-\right)^n  {\mathcal{O}} \right)\,.
\end{eqnarray}
Now, if we compute the $z$-expansion of $\mathcal{O}^\Omega$ we have
\begin{eqnarray}
\label{DDFL}
\mathcal{O}^\Omega_m = \int \frac{dz}{2\pi i} z^{m+1} \mathcal{O}^\Omega (z) = 
 \int \frac{dz}{2\pi i} \left(z + X^-(z, \bar z)\right) ^{m+1} \mathcal{O}(z) \,,
\end{eqnarray}
where we used integration by parts in the last step. This tells us that the dressing of an $h=1$ conformal primary operator $z$ expansion leads to different locality properties because $\mathcal{O}^\Omega$ is linked to the undressed one
by the change of coordinates
\begin{eqnarray}\label{coord-ch}
\mathcal{O}^\Omega(z, \bar z)= \partial_z w(z,\bar z)\mathcal{O}(w(z, \bar z))\,, \quad z= w +X^-(w, \bar w)\,.
\end{eqnarray}
Note that the above equation corresponds exactly to the transformation law for a holomorphic primary operator of dimension $h = 1$.
Similarly, by repeating the analysis with $X^- = 0$ and considering an antiholomorphic primary with $\bar{h} = 1$, we find an analogous result, with the transformation $\bar{z} = \bar{w} -X^+(w, \bar{w})$. \\

\paragraph{Dressed field in closed form.} The analysis of the previous paragraph suggests that, more generally, the dressing of an arbitrary conformal primary $\mathcal{O}(z,\bar z)$ can be obtained from its transformation rules under the following field-dependent coordinate transformation
\begin{eqnarray}\label{eq:trasfxpm}
     z= w + X^-(w,\bar w)\,, \quad  \bar{z} = \bar{w} + X^+(w, \bar{w}) \,.
\end{eqnarray}
Note, however, that the transformation \eqref{eq:trasfxpm} is not conformal. This implies that the finite transformation will be more complicated than the standard transformation of primaries under conformal maps. \\

We will now proceed to prove the ansatz \eqref{eq:trasfxpm} for primary operators with weights $h = \bar{h}$. These operators correspond to scalars of dimension $\Delta = h + \bar{h}$.
The transformation for spinning operators is a generalization of the following argument, and we will comment on it afterwards.
To prove the ansatz, we first bring the equation for the intertwiner into a standard Heisenberg evolution equation in an auxiliary parameter $T$, from $T = 0$ to $T = 1$, via the change of coordinates
$T=\exp(-t)$ in~\eqref{mass19}
  \beq
 {d\over dT} \Omega(T) =- \,\Omega(T) (\Sigma+ \bar \Sigma)\,, \quad \Omega(0)= \mathbb{I}\,,
 \eeq{prim1}
 whose solution is $\Omega|_{T=1}= \exp(\Sigma+\bar \Sigma)$. In addition, 
 we set $  \mathcal{O}_{T} \equiv \Omega^{-1}(T)\mathcal{O} \, \Omega(T) $ and $\mathcal{O}^\Omega = \mathcal{O}_{T=1}$. 
 The evolution equation in $T$ for any dressed operator is
 \beq
  {d\over dT}\mathcal{O}_T  =  -\Omega^{-1}(T)[\Sigma + \bar \Sigma,\mathcal{O}] \Omega(T)\,.
  \eeq{prim2}
When the operator $\mathcal{O}$ is a local field and also a conformal primary of the $c = c_{\text{CFT}}$ matter theory, the commutator $[\Sigma + \bar{\Sigma}, \mathcal{O}(z, \bar{z})]$ is the infinitesimal coordinate transformation with parameters $X^-(z, \bar{z})$ and $X^+(z, \bar{z})$, as shown in \eqref{eq:firstcommutator}. 
Using this result together with the definition of $\mathcal{O}_T(z, \bar{z})$ and equation \eqref{prim2}, we obtain
 \begin{equation}\label{prim3a}
 \begin{split}
 \frac{d}{dT}\mathcal{O}_T (z, \bar z) & =  - \Big( h\, \partial X_T^-\mathcal{O}_T + X_T^-\partial\mathcal{O}_T +\bar h\, \bar \partial X^+_T \mathcal{O}_T+X^+_T\bar \partial\mathcal{O}_T\Big)\,,\\
 \left. X_T^\pm(z,\bar z)\right. & \equiv \Omega^{-1}(T) X^\pm(z, \bar z) \Omega(T) \,.
  \end{split}
 \end{equation}
 We can now check whether our ansatz satisfies the differential equation \eqref{prim3a}. 
 Explicitly, it is
 \begin{equation}\label{eq:ansatzsolution}
     \begin{split}
       \mathcal{O}_T(z,\bar z)= \bigg(\!\det \frac{\partial (w,\bar w)}{\partial (z,\bar z)}\bigg)^{\frac{h+ \bar h}{2}}\,\mathcal{O}(w(z,\bar z,T),\bar w(z,\bar z,T))\,, \quad h=\bar h \,,
     \end{split}
 \end{equation}
 where $w(z,\bar z,T)$ and $ \bar w(z,\bar z,T)$ are the solutions of
 \begin{eqnarray}
        z= w +T X^-(w,\bar w)\,, \quad  \bar{z} = \bar{w} +TX^+(w, \bar{w}) \,.
 \end{eqnarray}
Let us first focus on finding $X^\pm_T(z, \bar{z})$. By differentiating the second line of \eqref{prim3a}, we obtain
\begin{equation}\label{eq:xpmdiffeq}
\begin{split}
     \frac{d}{dT}X^\pm_T(z,\bar z)&=-\big( X^-_T(z,\bar z)\partial X^\pm_T(z,\bar z)+ X^+_T(z,\bar z)\bar\partial X^\pm_T(z,\bar z)\big)\,, \\
     X^\pm_{T=0}(z,\bar z)&=X^\pm(z,\bar z)\,.
    \end{split}
\end{equation}
The solution is given by $X^\pm_T(z,\bar z)=X^\pm\big(w(z,\bar z,T),\bar w(z, \bar z ,T)\big)$, \emph{i.e.}~$X^\pm$ transform as scalars.
The initial condition is clearly satisfied.
Indeed, by differentiating this ansatz, we find
\begin{equation}\label{eq:xpmtimederivative}
    \frac{d}{dT}X^\pm_T(z,\bar z)=\partial_w X^\pm(w,\bar w)\partial_Tw+\bar \partial_{\bar w}X^\pm(w,\bar w) \partial_T\bar w\,.
\end{equation}
Moreover, by differentiating with respect to $T$ the identities $z = z\big(w(z, \bar{z}, T), \bar{w}(z, \bar{z}, T), T\big)$ and \\
$\bar{z} = \bar{z}\big(w(z, \bar{z}, T), \bar{w}(z, \bar{z}, T), T\big)$ at fixed $z$ and using the inverse Jacobian, we obtain
\begin{equation}
    \begin{pmatrix}
        \partial_T w \\
        \partial_T \bar w
    \end{pmatrix}
    =
   - \begin{pmatrix}
        \partial_z w & \bar\partial_{\bar z}w \\
        \partial_{z} \bar w & \bar \partial_{\bar z}\bar w
    \end{pmatrix}
    \begin{pmatrix}
        \partial_T z \\
        \partial_T \bar z
    \end{pmatrix}\,.
\end{equation}
Finally, by using $\partial_T z =  X^-(w, \bar{w}) = X^-_T(z, \bar{z})$ and $\partial_T \bar{z} =  X^+(w, \bar{w}) =  X^+_T(z, \bar{z})$, and inserting these into \eqref{eq:xpmtimederivative}, we recover \eqref{eq:xpmdiffeq}.\\

Now that we have shown that $X^\pm_T$ transform as scalars, we can repeat the analysis to show that the ansatz \eqref{eq:ansatzsolution} for the dressing of a generic primary $\mathcal{O}_T(z, \bar{z})$ satisfies the differential equation \eqref{prim3a}. 
This can be done by taking the derivative with respect to $T$ of \eqref{eq:ansatzsolution}, substituting $\partial_T w$ and $\partial_T \bar{w}$ as above, and then, with some involved but straightforward manipulations, reconstructing the first equation in \eqref{prim3a}.\\

At this point, we can take the transformation rule in \eqref{eq:ansatzsolution}, which we have just proven, and set $T = 1$. 
We then obtain a general solution for the dressed field of primary operators $\mathcal{O}(z, \bar{z})$ with conformal weights $h=\bar{h}$
\begin{equation}\label{eq:primarytransfnew}
\mathcal{O}^\Omega\!\left(w +X^-(w,\bar w), \bar w + X^+(w,\bar w)\right)\!=\! \bigg(\!\det\frac{\partial (z,\bar z)}{\partial(w,\bar w)}\bigg)^{\!\!-\frac{h+\bar h}{2}} \mathcal{O}(w,\bar w)\,, \quad h=\bar h\,.
   \end{equation}
Note that, as anticipated, even if $\mathcal{O}(z)$ does not depend on $\bar{z}$ and has $\bar{h} = 0$, the dressed operator $\mathcal{O}^\Omega(z, \bar{z})$ will depend non-trivially on $\bar{z}$.
As a consistency check, one can also expand \eqref{eq:primarytransfnew} to recover the nested commutators in the expansion \eqref{DDFA}.
Note that the deformed operator $\mathcal{O}^\Omega$ does not depend explicitly on the deformation parameter $\lambda$;  $\lambda$ appears in the correlators only through the OPE of the fields $X^\pm$.
\\

In the more general situation where the CFT operator $\mathcal{O}$ has arbitrary weights $h$ and $\bar h$, it becomes harder to guess an expression like \eqref{eq:ansatzsolution}, because it is not a priori obvious how a non-scalar operator should transform under a non-conformal transformation such as \eqref{eq:trasfxpm}.  
However, by computing the first commutators in the expansion \eqref{DDFA2}, one can still infer the correct ansatz, which can then be proven using methods similar to those employed in the case $h = \bar h$.  
The general formula is
\begin{equation}\label{eq:primarytransfgeneric}
    \mathcal{O}^\Omega(z,\bar z)=\bigg( \!\det \frac{\partial (w,\bar w)}{\partial (z,\bar z)} \bigg)^{\frac{h+\bar h}{2}} \!\big( \partial_z w \big)^{\frac{h-\bar h}{2}}\!\big( \bar \partial_{\bar z} \bar w \big)^{\frac{\bar h- h}{2}}\,\mathcal{O}(w,\bar w)\,,
\end{equation}
where we recall once again that $w(z,\bar z)$ and $\bar w(z,\bar z)$ are the solutions of
\begin{eqnarray}\label{eq:trasfxpm2}
     z= w + X^-(w,\bar w)\,, \quad  \bar{z} = \bar{w} + X^+(w, \bar{w}) \,.
\end{eqnarray}

In Section \ref{comp}, we will apply these explicit results to the analysis of deformed correlation functions.

   \subsection{Other vertex operators} \label{sec:othervert}
As discussed in Section \ref{sec:q0cohom}, this theory contains many vertex operators beyond the dressed CFT fields. In general, starting from an element of $H(\mathcal{Q}_0)$ with ghost number zero and definite weights $h$ and $\bar h$, one can repeat the analysis of Section \ref{dresscft}. 
The result is that the action of the intertwiner still corresponds to the field-dependent change of coordinates \eqref{eq:trasfxpm}, and that an expression analogous to \eqref{eq:primarytransfgeneric} continues to hold.
For elements with non-zero ghost number, one would need to generalize the calculations.
\\

As an example, 
we can consider the vertex operators discussed in \cite{Callebaut:2019omt}, namely
\begin{eqnarray}
    \label{opA}
    V_{\mathcal O}  = c\bar c \, \mathcal{O}_\Delta e^{i p \cdot X}\,,
\end{eqnarray}
where $p \cdot X = p^{+} X^{-} + p^{-} X^{+}$ and $\mathcal{O}_\Delta$ is a CFT operator of conformal dimension $\Delta$.
The operator $V_{\mathcal O}$ belongs to the cohomology $H(\mathcal{Q}_0)$ in \eqref{holC}.  
Indeed, we may choose the representative
\begin{equation}
    U_\mathcal{O} = \mathcal{O}_\Delta\, c \bar c :\!\delta(X^{-})\, \delta(X^{+})\!: \ ,
\end{equation}
and rewrite the Dirac delta functions using their integral representations as follows (see also \cite{Aharony:2023dod})
\begin{eqnarray}
    \label{opAB}
    U_\mathcal{O} =  \int_{-\infty}^\infty dp^+ \int_{-\infty}^\infty dp^- \, {\mathcal O}_\Delta c\bar c
     :\!e^{i p \cdot X}\!:\,.
\end{eqnarray}
The integrand 
coincides with the vertex operator \eqref{opA}.  
In order to show that the integrand can be written in terms of the interwiner, namely as 
\begin{equation}
    {\mathcal M} = :e^{-(\Sigma +\bar \Sigma)} V_{\mathcal O}  e^{\Sigma + \bar \Sigma}:\,,
\end{equation}
{with $\mathcal M$ in $H(\mathcal{Q})$},
 we first observe that it is $Q_0$- and $\bar Q_0$-closed for any values of $\Delta$, $p^{+}$, and $p^{-}$, as can be easily checked.
For $p^{\pm} \neq 0$, we can express it as follows
\begin{eqnarray}
    \label{opAC}
     V_{\mathcal O} = \Big\{Q_0, \Big[\bar Q_0,\frac{{\mathcal O}_\Delta}{p^+p^- } :\!e^{i p \cdot X}\!:\Big]\Big\}\,.
\end{eqnarray}
However, we cannot regard $V_\mathcal{O}$ as exact, since it is the variation of an object that is not locally integrable in momentum space.\footnote{
Note that this is precisely the condition for its Fourier transform to be a compactly supported distribution in position space.
}
In the special case $\Delta - \lambda\, p^{+} p^{-} - 1 = 0$, the vertex operator $\mathcal{O}_\Delta e^{i p \cdot X}$ has total conformal weights $h = 1$ and $\bar h = 1$. 
In this situation, using {$\mathcal{Q}_0 e^{\Sigma+\bar\Sigma} =e^{\Sigma+\bar\Sigma} \mathcal{Q}$}, we find that  
\begin{eqnarray}
    \label{nPAGa}
  {\mathcal M}(p^+, p^-)  &=& 
  \Big\{Q,\Big[\bar Q, 
   \frac{1}{p^+ p^-}
  :e^{-(\Sigma + \bar \Sigma)} \Big( 
 e^{i p\cdot X} {\mathcal O}_\Delta\Big)
  e^{\Sigma + \bar \Sigma}: \Big]\Big\} 
  \nonumber \\ 
  &=& 
   \Big\{Q,\Big[\bar Q, 
   \frac{1}{p^+ p^-}
  e^{-i (\partial_{p^-} \partial +  \partial_{p^+} \bar\partial)}
 :e^{i p\cdot X} {\mathcal O}_\Delta:
 \Big]\Big\}\,.
\end{eqnarray} 
The 
differential operator in the exponent is the {Laplacian} in the space of 
coordinates $(p^-,z)$ and $(p^+,\bar z)$ acting on the vertex operator. Working out this expression is not very illuminating, but we can compute the invese Fourier transformation of \eqref{opAB} 
integrating ${\mathcal M}(p^+, p^-) $ over 
$p^\pm$ get the result
\begin{eqnarray}
    \label{nPAGb}
   \int dp^+dp^-   {\mathcal M}(p^+, p^-)  = 
   \Big\{Q,\Big[\bar Q, \Theta(X^- + \partial) \Theta(X^+ + \bar\partial) {\mathcal O}_\Delta
   \Big]\Big\} \,,
\end{eqnarray}
where it shows that the vertex operator indeed belongs to the cohomology of $\mathcal{Q}$, since it is the BRST variation of an object that lies outside the allowed functional space (compactly supported distributions). The resulting expression is a power series of derivatives of delta functions of $X^\pm$, and therefore a linear combination of the vertices discussed in the previous section.

\section{Deformed Correlators}\label{comp}

In this section, we analyze how the intertwiner formalism developed in the previous sections can be used to \emph{define} deformed correlation functions, and then applied to the derivation of flow equations, the explicit computation of deformed correlators, and the derivation of conservation equations.
  \\


Given the local observables $\mathcal{O}(z, \bar z)$, we aim to define their deformed correlation functions for generic values of the deformation parameter $\lambda$. 
In our construction, the natural objects to consider are gauge-invariant correlators of the dressed observables, which are in one-to-one correspondence with the CFT local operators.
These gauge-invariant correlators are defined as expectation values of elements of the BRST cohomology $H(\mathcal{Q})$ in the $SL(2,\mathbb{C})$-invariant vacuum $|0\rangle$. 
The vacuum $|0\rangle$ is annihilated by all modes $a^\pm_n$ of the $X^\pm$ fields for $n \geq 1$, by the ghost modes $b_n$ with $n \geq -1$ and $c_n$ with $n \geq 2$, and by the Virasoro generators $L_n$ of the original CFT for $n \geq -1$.\footnote{
For a holomorphic operator $\mathcal{O}$ of weight $h$, the mode expansion is $
\mathcal{O}(z)=\sum_{n \in \mathbb{Z}} \mathcal{O}_n\, z^{-h-n}\,,$
and similarly for antiholomorphic operators.
}
The vacuum also satisfies the standard BPZ normalization condition 
$\langle 0| c_{-1} c_0 c_1 \bar c_{-1}\bar c_0 \bar c_1|0\rangle = 1$.
To obtain a non-zero correlator between the dressed observables 
$\mathcal{O}^\Omega_i(z_i,\bar z_i)$ that contain only matter fields, 
we must insert three gauge-fixing operators $c\bar c$. 
As usual, we can place these insertions at the points $0$, $1$, and $\infty$.\\

We define the deformed correlators $\langle \,\cdots\, \rangle_{\lambda}$ as follows
\begin{equation}\label{OBE}
    \langle \mathcal{O}_1(z_1,\bar z_1)\dots \mathcal{O}_n(z_n,\bar z_n)\rangle_{\lambda}
    \;\equiv\;
    \langle 0 | \,c \bar c(0) \,c\bar c(1) \, c\bar c(\infty)\, \mathcal{O}_1^{\Omega}(z_1,\bar z_1)\dots \mathcal{O}_n^{\Omega}(z_n,\bar z_n)\, |0\rangle \, .
\end{equation}
First of all, we need to ensure that the correlator on the right-hand side of \eqref{OBE} is indeed gauge invariant.
This is guaranteed by the fact that we insert only operators that are elements of the cohomology of the full BRST charge $\mathcal{Q} = \mathcal{Q}_{0} + \mathcal{Q}_{1}$.
It is also straightforward to verify directly that the dressed observables $\mathcal{O}^\Omega_{i}(z_i, \bar z_i)$ defined in \eqref{DDFA} commute with the BRST charge, since
\begin{equation}
    [\mathcal{Q}, \mathcal{O}_i^\Omega(z_i,\bar z_i)]=\Omega^{-1}[\mathcal{Q}_0,\mathcal{O}(z_i,\bar z_i)]\Omega=0\,.
\end{equation}
Therefore, if we insert the BRST charge $\mathcal{Q}$ anywhere in the correlator on the right-hand side of \eqref{OBE}, we can always move it to the far right, where it annihilates the vacuum $|0\rangle$. 
Note that, since {no operator $\mathcal{O}^\Omega_{i}(z_i, \bar z_i)$  contains} ghost insertions, the ghost sector factorizes, and we can simply evaluate the correlator on the matter vacuum $|0\rangle_m$\footnote{More precisely, this vacuum is 
$|0\rangle_m \equiv |0\rangle_{\text{CFT}} \otimes |0\rangle_{X}$, 
where $|0\rangle_{\text{CFT}}$ is the original CFT vacuum, and 
$|0\rangle_{X}$ is the vacuum annihilated by all modes $X^\pm_n$ for $n \geq 1$. 
}
\begin{equation}\label{OBE2}
    \langle \mathcal{O}_1(z_1,\bar z_1)\dots \mathcal{O}_n(z_n,\bar z_n)\rangle_{\lambda}
    \;\equiv\;
    _m \!\langle 0 |  \mathcal{O}_1^{\Omega}(z_1,\bar z_1)\dots \mathcal{O}_n^{\Omega}(z_n,\bar z_n)\, |0\rangle_m \, .
\end{equation}
The first condition for our definition to be consistent is that, in the limit $\lambda \to 0$, we should recover the undeformed CFT correlators.
To see this, first note that the deformation parameter appears only in the OPE between the fields $X^\pm$. Therefore, the expansion in $\lambda$ is equivalent to an expansion in powers of the fields $X^\pm$. 
In particular, for an arbitrary primary, by expanding \eqref{eq:primarytransfgeneric} we can write
\begin{equation}\label{eq:firstexpansion}
\mathcal{O}^\Omega_{\Delta}(z,\bar z)
   = \mathcal{O}_\Delta(z,\bar z) + \text{O}(X^\pm)\,,
\end{equation}
and the omitted terms contribute to the deformed correlators only through terms that vanish in the limit $\lambda \to 0$.
It then follows immediately that, in this limit, we recover the undeformed CFT correlators.
It is also interesting to understand how Poincaré invariance of the deformed correlators defined in \eqref{OBE2} is ensured.
The action \eqref{caA} and the Virasoro constraints \eqref{cab} are both invariant under a target-space symmetry group $ISO(1,1)$, generated by translations $X^\pm \to X^\pm + a^\pm$ and Lorentz boosts $X^\pm \to e^{\pm \alpha} X^\pm$.
Taking into account the redefinition \eqref{etCC}, the effect of these symmetries on the fields $X^\pm$ is
\begin{equation}
    X^\pm(w,\bar w) \to X^\pm(w,\bar w) + a^\pm \,, \quad X^\pm(w,\bar w) \to  e^{\pm \alpha}X^\pm(w,\bar w)+(e^{\pm\alpha}-1)w^\pm\,,
\end{equation}
where we used the compact notation $w^+=w$ and $w^-=\bar w$.
By looking at Equation~\eqref{eq:trasfxpm2}, it immediately follows that the net effect of this internal symmetry on the deformed operators $\mathcal{O}^\Omega(z,\bar z)$ is to perform the inverse of the corresponding isometry on the coordinates $z$ and $\bar z$.
Thus, the internal symmetry of the fields $X^\pm$ becomes a spacetime symmetry for the deformed operators.\\
We can now move on to analyze some of the consequences that follow from the definition \eqref{OBE}.
\\


First, we can use the definition \eqref{OBE} to compute the variation of correlators with respect to the deformation parameter. This yields a \emph{flow equation}, analogous to the one obtained in \cite{Cardy:2019qao}.  
By taking the derivative of \eqref{OBE} with respect to the deformation parameter, we get
\begin{eqnarray}
    \label{OBD}
     \frac{d}{d\lambda}
   \Big\langle \prod_i \mathcal{O}_i(z_i, \bar z_i)\Big\rangle_{\lambda} = \sum_i
    {}_m\!\langle 0|
    \frac{d}{d\lambda} \mathcal{O}_i^\Omega(z_i, \bar z_i)
    \prod_{j\neq i} \mathcal{O}_j^\Omega(z_j, \bar z_j)|0\rangle_m \, ,
\end{eqnarray}
where $\mathcal{O}_i^\Omega$ satisfies a suitable flow equation. 
Therefore, the variations of a correlator can be translated into variations 
of the inserted observables, as in \cite{Cardy:2019qao}.
Now, if we consider the dressed observables 
${\mathcal O}_i^\Omega$, we 
find that 
\begin{eqnarray}
    \label{OBEE}
     \frac{d}{d\lambda} \mathcal{O}^\Omega(z,\bar z) &=&  \frac{d}{d\lambda}\left( \Omega^{-1} {\mathcal O}(z,\bar z) \Omega  \right) = 
 \Big[ \mathcal{O}^\Omega(z,\bar z), \big(\Omega ^{-1}  \frac{d}{d\lambda} \Omega  \big)\Big]\,,
\end{eqnarray}
and using the defining equations of the intertwiners, we have 
\begin{eqnarray}
    \label{OBF}
     \frac{d}{d\lambda} \mathcal{O}^\Omega(z,\bar z) =  
    -\Big[\frac{d}{d \lambda}\big( \Sigma + \bar \Sigma \big) ,\mathcal{O}^\Omega(z,\bar z) \Big]\,,
\end{eqnarray}
which is the flow equation for the dressed observables. \\
Using the explicit equations for $\Sigma$ and 
$\bar \Sigma$ {given in}~\eqref{RD}, we finally get the analogue of Cardy's formula \cite{Cardy:2019qao}\,\footnote{
To get the correct result, one should take into account that the fields $X^\pm$ depend on $\lambda$, since it appears in their OPE \eqref{eq:OPE1}. 
Hence, one should define the rescaled fields $\hat X^\pm \equiv X^\pm / \sqrt{\lambda}$, compute the derivative, and then express it in terms of the original $X^\pm$.}
\begin{equation}\label{OBG}
    \begin{split}
         & \frac{d}{d\lambda} \mathcal{O}^\Omega(z, \bar z) =  \\
  & =- \frac{1}{2\lambda^2} 
      \!\left(\int \!\!\frac{dw}{2 \pi i} (\partial X^+  \partial X^-)X^-(w,\bar w) \right. - \left.
     \int\!\! \frac{d\bar w}{2 \pi i} (\bar \partial X^+ \bar \partial X^-) X^+(w,\bar w) \right) \!\mathcal{O}^\Omega(z, \bar z) \,,
    \end{split}
\end{equation} 
where it is understood that the integrals should be interpreted as in \eqref{eq:integralsdz}. 
It follows immediately from \eqref{OBF} that the variation $\frac{d}{d\lambda}\mathcal{O}^\Omega(z,\bar z)$ acts as a derivation on the algebra of observables, as expected from the automorphisms generated by the intertwiner $\Omega $. This confirms the arguments of Cardy in \cite{Cardy:2019qao}. 
Notice that the algebra automorphism~\eqref{OBG} exists even if the action of $\Sigma$ and $\bar{\Sigma}$ on the vacuum is not well defined due to the presence of underived $X^\pm$ operators.
 \\

We can also apply the relation \eqref{eq:primarytransfnew} to explicitly compute correlators, in order to see how it reproduces known results. 
For example, we can compute the deformed two-point function of a scalar operator $\mathcal{O}_\Delta(z,\bar z)$, which has conformal weights $h=\bar h=\Delta/2$. 
To do this, we expand as in \eqref{eq:firstexpansion} and keep track of the higher–order terms. 
At first order in $\lambda$, we can have only a single OPE, thus we expand $\mathcal{O}^\Omega_\Delta$ to linear order in the fields $X^\pm$
\begin{eqnarray}\label{eq:primaryscalarexpansion}
    \mathcal{O}^\Omega_{\Delta}(z,\bar z)= \mathcal{O}_\Delta(z,\bar z) -\Big( X^-\partial \mathcal{O}_\Delta+X^+\bar \partial\mathcal{O}_\Delta+\frac{\Delta}{2}\big( \partial X^- \mathcal{O}_\Delta+ \bar \partial X^+\mathcal{O}_\Delta\big)\! \Big)+\text{O}\big((X^\pm)^2\big)\,.
\end{eqnarray}
Now, we can compute the deformed two-point function
\begin{eqnarray}
    \langle \mathcal{O}^\Omega_\Delta(w,\bar w)\mathcal{O}^\Omega_\Delta(z,\bar z)\rangle_\lambda\,.
\end{eqnarray}
By using the expansion \eqref{eq:primaryscalarexpansion} and taking all the possible {contractions}, we find
\begin{equation}
    \begin{split}
        \langle \mathcal{O}^\Omega_\Delta(w,\bar w)\mathcal{O}^\Omega_\Delta(z,\bar z)\rangle_\lambda & =\langle\mathcal{O}_\Delta\mathcal{O}_\Delta\rangle_{CFT} -\lambda \bigg(\!\log(\kappa |w - z|)(\partial_w \bar \partial_{\bar z}+\bar \partial_{\bar w}\partial_z) \\
        & -\frac{\Delta}{2}\Big(\,\frac{\partial_w}{\bar w-\bar z}+\frac{\bar \partial_{\bar w}}{ w- z}-\frac{\partial_z}{\bar w-\bar z}-\frac{\bar \partial_{\bar z}}{ w- z}\,\Big)
        \!\bigg)\langle \mathcal{O}_\Delta  \mathcal{O}_\Delta \rangle_{CFT} + \text{O}(\lambda^2)\,.
    \end{split}
\end{equation}
More explicitly, this is
\begin{equation}
   \langle \mathcal{O}^\Omega_\Delta(w,\bar w)\mathcal{O}^\Omega_\Delta(z,\bar z)\rangle_\lambda  = \frac{1}{|w-z|^{2\Delta}}+ \frac{2 \lambda \Delta}{|w-z|^{2\Delta+2}}\Big( \Delta \log(\kappa |w-z|)-1\Big)\,,
\end{equation}
where the last constant can be reabsorbed into a redefinition of $\kappa$. 
This result is in agreement with \cite{Kraus:2018xrn} and \cite{Cardy:2019qao}.
\\



Finally, we can use the relation \eqref{eq:primarytransfgeneric} to analyze the behavior of conserved currents under the deformation.
Consider a conserved holomorphic current $\mathcal{J}(z)$. 
This current has conformal weights $h = 1$, $\bar{h} = 0$, and the corresponding deformed operator is given by
\begin{eqnarray}
  \begin{split}
      &  \mathcal{J}^\Omega \!\left(w +  X^-(w,\bar w), \bar w +X^+(w,\bar w)\right)= \mathcal{N}_\mathcal{J}(w,\bar w)\, \mathcal{J}(w)\,, \\
    & \mathcal{N}_\mathcal{J}(w,\bar w) \equiv \sqrt{\partial_z w\bigg(\partial_zw-\frac{\partial_z \bar w \bar \partial_{\bar z}w     }{\bar \partial_{\bar z}\bar w}\bigg)}\,.
  \end{split}
\end{eqnarray}
The right-hand side of the above equation depends on $\bar{w}$ only through the prefactor $\mathcal{N}_J(w,\bar w)$, which leads to a non-trivial identity for the deformed operator
\begin{eqnarray}\label{eq:deformedconservation}
   \bar D_{\bar w} \Big( \mathcal{J}^\Omega \!\left(w +  X^-(w,\bar w), \bar w + X^+(w,\bar w)\right)\Big)=0\,,
\end{eqnarray}
where we used the covariant derivative $\bar D_{\bar w}$ defined as
\begin{equation}
  \bar D_{\bar w} \mathcal{O}(w,\bar w) \equiv \bar \partial_{\bar w} \Big(  \mathcal{N}_\mathcal{J}(w,\bar w)^{-1} \mathcal{O}(w,\bar w) \Big)\,.
\end{equation}
Equation \eqref{eq:deformedconservation} should be understood as the deformed version of the conservation equation.

\section{$J\bar J$ deformations}\label{jjcohom}

In this section, we briefly discuss the intertwiner formalism for the $J\bar J$ deformations (for a review, we refer to \cite{Borsato:2023dis}). We assume an action $S_0[\Phi]$ invariant under a $U(1)\times U(1)$ symmetry. The symmetry
 can be gauged by promoting the action $S_0[\Phi|B]$ to be gauge invariant 
\begin{eqnarray}
    \label{JJA}
    S_0[ e^{\alpha^a q_a} \Phi| B^a + d \alpha^a ] = 
    S_0[\Phi|B] \,,
\end{eqnarray}
where we introduced the gauge fields $B^a$. 
{To deform} the theory, we follow \cite{Dubovsky:2023lza} and we add 
{a $\lambda$-dependent} term 
\begin{eqnarray}
    \label{JJB}
    S = S_0[\Phi|B] + \lambda \int d^2z \, \epsilon_{ab}
    (\partial X^a - B^a) (\bar \partial X^b - \bar B^b) \,.
\end{eqnarray}
The equations of motion are 
\begin{eqnarray}
    \label{JJC}
    &&\partial (\bar \partial X^a - \bar B^a) - \bar \partial (\partial X^a - \bar B^a) = - \partial \bar B^a + \bar \partial B^a =0 ,\nonumber \\ 
    && \bar \partial X^a - \bar B^a + \bar J^a =0 , \nonumber \\
&&  - (\partial X^a - B^a) + J^a =0 \,,
\end{eqnarray}
where $\bar J^a = \epsilon^{ab} \frac{\delta S_0}{\delta B^b(x)}$ and $J^a = \epsilon^{ab} \frac{\delta S_0}{\delta \bar B^b(x)}$ are the {$U(1)\times U(1)$} currents {computed from the action $S_0$.} Note that $X^a$ drops out of
 the first equation, which can also be written as 
$d B^a =0$. The equations of motion are invariant under the classical BRST symmetry 
\begin{eqnarray}
    \label{JJD}
    s\, B^a = \partial c^a\,, ~~~~~ 
    s\, \bar B^a = \bar \partial c^a\,, ~~~~~
    s\, X^a = c^a\,.
\end{eqnarray}
The field $X^a$ plays the role of the St\"uckelberg field. Equations \eqref{JJD} are consistent with the current conservation law $\partial \bar J^a - \bar \partial J^a =0$. By solving the second and third equations and plugging them back into the action, we get the deformed action 
\begin{eqnarray}
    \label{JJE}
 S = S_0[\Phi|B] - 
 \lambda \epsilon_{ab} \int d^2z  J^a \bar J^b  \,  .
\end{eqnarray}
To quantize the system {before solving the constraints} we need a gauge fixing. This can be done by requiring $d \star B^a =0$, which in terms of complex fields means $\partial \bar B^a + \bar \partial B^a =0$. \\

Rewriting the  Lagrangian with the couplings to the $B$-fields, we have 
\begin{eqnarray}
    \label{newJJA}
    {\mathcal L} = B_a \left(\bar J^a - \frac12\bar\partial \rho^a - \epsilon^{ab} \bar \partial X_b \right) + 
    \bar B_a \left(- J^a - \frac12\partial \rho^a + \epsilon^{ab} \partial X_b \right) + \epsilon^{ab} B_a \bar B_b - b_a \partial \bar \partial c^a\,,
\end{eqnarray}
where $\rho$ is a Lagrange multiplier that enforces the gauge-fixing condition.
Solving for $B_a$ and $\bar B_a$ and inserting back into the Lagrangian we have following 
\begin{eqnarray}
    \label{newJJB}
    {\mathcal L} = \epsilon_{ab} J^a \bar J^b - J^a \epsilon_{ab} \bar\partial Y^{+,b} + \partial Y^{-,a}\epsilon_{ab} \bar J^b - 
    \partial Y^{-,a} \epsilon_{ab} \bar \partial Y^{+,b}
    - b_a \partial \bar \partial c^a\,,
\end{eqnarray}
where we have defined combinations $Y^{+,a} \equiv \frac12 \rho^a + \epsilon^{ab} X_b$ and $Y^{-,a} =\frac12 \rho^a - \epsilon^{ab} X_b$
which transfom as follows 
\begin{eqnarray}
    \label{newJJC}
    s \, Y^{+,a} = \epsilon^{ab} c_b\,, ~~~~~~~
    s \, Y^{-,a} = -\epsilon^{ab} c_b\,, ~~~~~~~
    s b^a = Y^{+,a}  + Y^{-,a}\,, ~~~~~ s \rho^a =0\,, ~~~~~s c^a =0\,.
\end{eqnarray}
The action is invariant due to the conservation of $J$ currents. Note that $\rho^a =  Y^{+,a}  + Y^{-,a}$ which is clearly invariant under  \eqref{newJJC}. {We compute the BRST charge by separating into a left and a right part as follows}
\begin{eqnarray}
    \label{newJJD}
    Q_L = \int \left( c_a \epsilon^{ab} \Pi^{+}_b + Y^{-, a} 
    \Pi^b_a + c_a J^a\right)\,, ~~~~~~
    Q_R =\int \left(- c_a \epsilon^{ab} \Pi^{-}_b + Y^{+, a} \Pi^b_a + c_a \bar J^A\right)\,, 
\end{eqnarray}
where $\Pi^{\pm}_b$ are the conjugated momenta to $Y^{\pm,a}$ 
and $\Pi^b_a$ is the conjugate momentum of $b^a$.
Both charges are nilpotent $Q^2_L = Q^2_R =0$ and they anticommute 
$\{Q_L, Q_R\}=0$. Note that the first two terms of the BRST charges are the $0$-charge pieces, the last terms in both equations 
are charge $+1$ pieces. Therefore, the intertwiner $\Omega = e^{\Sigma + \bar \Sigma}$ interpolates between the $0$-charge pieces and the complete ones. {The intertwiner is}  obtained by using the 
$R$-operator
\begin{eqnarray}
    \label{newJJE}
    R_L = \int b_a \Pi^{-}_a  + \Pi^c_a \epsilon^{ab}Y^{+}_b  \,, ~~~~~
R_R = \int b_a \Pi^{+}_a  - \Pi^c_a  \epsilon^{ab} Y^{-}_b\,,
\end{eqnarray}
where $\Pi^c_a$ is the conjugate momentum of $c_a$. Computing the 
anticommutators {of} $R_{L/R}$ with the BRST charges $Q_{L/R}$ we have 
\begin{eqnarray}
    \label{newJJEA}
    \{Q_L, R_L\} = i S_L\,, ~~~~~~~~
    \{Q_R, R_R\} = i S_R\,, ~~~~~~~~
    \{Q_L, R_R\} = \{Q_R, R_L\} =0\,.
\end{eqnarray}
where $S_{L/R}$ are the counting operators that assign the charges to the fields of the theory. The last equations {show the mutual independence} of the two sectors. 
Finally, computing the anticommutators with the $1$-charge pieces we get 
\begin{eqnarray}
    \label{newJJF}
    \Sigma_L = \Big\{ \int c_a J^a, R_L\Big\} = \int J_a \epsilon^{ab} Y^{+}_b \,, 
~~~~
 \Sigma_R = \Big\{ \int c_a \bar J^a, R_R\Big\} = \int \bar J_a 
 \epsilon^{ab} Y^{-}_b \,, 
\end{eqnarray}
The intertwiner is given by $\Omega = e^{ \Sigma + \bar \Sigma }$, where $\Sigma + \bar \Sigma = \int \left( J_a \epsilon^{ab} Y^{+}_b  + \bar J^a\epsilon^{ab} Y^{-}_b \right)$.\\

To compute the action of the intertwiner on the spectrum, we can look at the OPE of the full energy-momentum tensor $T = T_{CFT} + T_{Y} + T_{bc}$ on an intertwined operator 
${\mathcal O}^\Omega = \Omega^{-1} {\mathcal O} \Omega$. 
Note that {the} auxiliary $Y$ and ghost systems do not change the total central charge, since they form a topological quartet system with central charge $c = \pm 2$. 
We find
\begin{eqnarray}
     \label{newJJG}
     T(z) {\mathcal O}^\Omega(w) = \Omega^{-1} \Big( \Omega T \Omega^{-1}  {\mathcal O}(w) \Big) \Omega\,.
\end{eqnarray}
Moreover, we have
\begin{equation}
\label{eq:JJbarshift}
    \Omega T_{Y} \Omega^{-1} =  T_{Y}  
+ \bar J^a \epsilon_{ab} \partial Y^{-,b} 
+ J^a \epsilon_{ab} \bar\partial Y^{+,b} + \epsilon_{ab} J^a \bar J^b\,.
\end{equation}
Therefore, it follows that the energy-momentum tensor is shifted by the charges of the operator ${\mathcal O}$.

To build other observables of the deformed theory, we apply the intertwiner $\Omega$ to a charged field 
$\mathcal{A}(\Phi(x)) $. We find
\begin{eqnarray}
    \label{tcC}
    {\mathcal O}_\mathcal{A}(x)
    = \Omega \mathcal{A}(\Phi(x)) \Omega^{-1} 
    = e^{i q^a \epsilon_{ab} X^b + \dots} \mathcal{A}(\Phi(x))\, .
\end{eqnarray}
Here the first term in the exponent reproduces the expression given in equation~(2.7) of ref~\cite{Dubovsky:2023lza}. The additional terms come from the nonzero mode of $X^a$. The observable ${\mathcal O}_\mathcal{A}(x)$ is the gauge-invariant observable constructed
 in terms of the $\mathcal{A}(\Phi(x))$.  
 
 Finally, gauge invariant correlators can be constructed in a similar way as in Section \ref{comp}. 


\section{Conclusions}\label{sec:conclusions}

In this work we analyzed the $T\bar T$ deformation using its formulation as a CFT coupled to two-dimensional dynamical gravity. 
The main novelty of our approach consists in applying BRST quantization together with the intertwiner method developed in~\cite{Grassi:2024vkb}. 
Within this framework, we showed that one can explicitly identify physical observables of the $T\bar T$--deformed theory and provide a non-perturbative definition of their deformed correlators in terms of worldsheet correlation functions. 
Furthermore, we showed that these deformed correlators can be expanded for small values of the deformation parameter and {can be} directly matched to those obtained in a perturbative QFT setting. \\

The importance of this result is twofold. 
On the one hand, we believe that our construction provides a setting in which sharp questions about quantum aspects of the $T\bar T$ deformation beyond perturbation theory can be formulated, studied, and eventually answered. 
One notable example is the precise realization of the idea that the $T \bar T$ deformation can be obtained from the undeformed CFT through a field-dependent change of coordinates. This is made explicit in equations \eqref{eq:primarytransfnew} and \eqref{eq:primarytransfgeneric}. 
As already emphasized in the Introduction, exploring the non-perturbative regime would be of considerable interest, and our construction provides a natural starting point for such an exploration. \\
On the other hand, the formalism also offers practical advantages. 
In particular, the deformed correlators introduced in Section~\ref{comp} are expressed entirely in terms of undeformed CFT correlators together with the free correlators of the additional fields $X^\pm$.
The complexity of the deformation is thus fully encoded in the definition of the deformed observables $\mathcal{O}^\Omega$, which has been largely worked out using the intertwiner method in Section~\ref{dresscft}. 
In particular, our method could also be useful to push the perturbative analysis of correlation functions to higher orders, as it does not require integrations or elaborate regularisation procedures, but only an expansion of the known deformed observables~$\mathcal{O}^\Omega$ together with simple Wick--like contraction rules. \\

We would like to further emphasize one point: definitions of deformed observables that are similar, at least in spirit, to our dressed observables in \eqref{eq:primarytransfnew} and \eqref{eq:primarytransfgeneric}, have already appeared in the recent literature.
Indeed, two relevant examples are provided by \cite{Hirano:2025alr}, which uses the massive gravity approach, and \cite{Hirano:2024eab}, which employs the dynamical coordinate formalism.
In both cases, as in our work, the deformation is implemented through a field-dependent coordinate transformation $(z,\bar z)\!\to\!(w,\bar w)$. The deformed observable is then given by the undeformed CFT operator evaluated at the transformed coordinates, $\mathcal{O}\big(w(z,\bar z),\bar w(z,\bar z)\big )$, possibly multiplied by a prefactor depending only on derivatives of $(w,\bar w)$.
A remarkable novel feature of our framework is that the form of the deformed observables is entirely fixed by a gauge-invariance principle.
Indeed, as discussed in Section~\ref{comp}, in our description conformal symmetry is a gauge symmetry, while the Poincaré symmetry of the deformed theory arises from an internal symmetry acting on the additional fields $X^\pm$.
This means that, for the deformed observable $\mathcal{O}^\Omega(z,\bar z)$ to be physical, it must be invariant under conformal transformations $(z,\bar z)\!\to\!(z+\xi(z),\bar z+\bar\xi(\bar z))$. 
This requirement completely fixes $\mathcal{O}^\Omega(z,\bar z)$ in terms of the undeformed operator $\mathcal{O}$, the additional fields $X^\pm$, and their descendants. \\

The natural next step is to apply the formalism developed here to reproduce and then generalize other known structural results about the $T\bar T$ deformation. Further developments along these lines are in progress and will be presented in future work. One concrete target is to reproduce the all--order expressions for the leading logarithms of deformed two-- and three--point functions obtained in \cite{Hirano:2025alr}, and to examine whether our method can be used to further generalize these results. Another direction is to construct, within our framework, the infinite set of symmetries discussed in \cite{Guica:2022gts}. It would also be worthwhile to clarify the role of the other gauge--invariant observables such as the vertex operators briefly discussed in Section~\ref{sec:othervert}, and to understand how they relate to the ``target--space'' operators analyzed in \cite{Aharony:2023dod}. Finally, in the long term, it would be interesting to explore whether the framework developed here can serve as a foundation for a bootstrap--type program for $T\bar T$--deformed theories.
A precise, non--perturbative definition of deformed correlators is only the first necessary step toward such a program, and many additional structural constraints would be required to even formulate it in a well-defined way.

 \subsection*{Acknowledgements} 
 We would like to thank Victor Gorbenko, Monica Guica, Ruben Monten, Stefano Negro, Shota Komatsu and Xi Yin for useful discussions.
 EdS's research is partially supported by the MUR PRIN contract 2020KR4KN2, ``String
Theory as a bridge between Gauge Theories and Quantum Gravity'', and by the INFN project 
ST\&FI, ``String Theory \& Fundamental Interactions.''
M.P. is supported in part by NSF grant PHY-2210349. M.P. and P.A.G. would like to thank the CERN TH Department for its hospitality during the
completion of this work. P.A.G. is supported by HORIZON-MSCA-2021-SE-01-101086123 CaLIGOLA.

\appendix

\section{The intertwiner for a BRST operator with an uncharged term}\label{app:intzerocharge}
In this section, we briefly discuss {the case} when the exact BRST charge $\mathcal{Q}^B$ contains a term with vanishing $S$-charge
\begin{equation}
    \mathcal{Q}^B=\mathcal{Q}_0 + \hat{\mathcal{Q}}_0 +\!\!\sum_{0<n<N}\!\! \mathcal{Q}_n \, , 
\end{equation}
where now the counting operator $S=\{\mathcal{Q}_0,\mathcal{R}\}$ satisfies $[S,\mathcal{Q}_n]=n\mathcal{Q}^n$ and $[S,\hat{\mathcal{Q}}_0]=0$.
We can still define the charge
\begin{equation}
    \mathcal{Q}(t)=\mathcal{Q}_0+\mathcal{Q}_I(t)=\mathcal{Q}_0+\hat{\mathcal{Q}}_0+ \!\!\sum_{0<n<N} \!\! e^{-nt}  \mathcal{Q}_n\,, 
\end{equation}
which now interpolates between
$\mathcal{Q}(0) = \mathcal{Q}^B$ at $t = 0$ and $\mathcal{Q}(+\infty) = \mathcal{Q}_0 +\hat{\mathcal{Q}}_0$ at $t = +\infty$.
All the arguments in Section \ref{int1} can still be applied, with the only difference that the intertwiner operator $\Omega(t)$ now gives
\begin{equation}
    \mathcal{Q}_0+\hat{\mathcal{Q}}_0  =\Omega(0) \mathcal{Q}^B \Omega^{-1}(0)\,, 
\end{equation}
since the boundary condition is $\mathcal{Q}(t\!=\!+\infty)=\mathcal{Q}_0+\hat{\mathcal{Q}}_0$. \\

If $\lim_{t\rightarrow+\infty} \{ \mathcal{Q}_I(t),\mathcal{R} \} = \{ \mathcal{Q}^0,\mathcal{R} \}$ vanishes, then the $t$-evolution of any operator $\mathcal{O}$, defined as $\mathcal{O}(t)=\Omega(t)\mathcal{O}\Omega^{-1}(t)$, admits a well-defined limit as $t \rightarrow +\infty$.  
Even if $\{ \mathcal{Q}^0, \mathcal{R} \}$ does not vanish, the operator $\mathcal{Q}(t)$ still admits a well-defined limit as $t \rightarrow +\infty$, since from the condition $(\mathcal{Q}_0 + \hat{\mathcal{Q}}_0)^2 = 0$ we find that
\bea
\lim_{t\rightarrow+\infty}[ \{ \mathcal{Q}_I(t),\mathcal{R} \}  ,\mathcal{Q}(t)] &=& [ \{ \hat{\mathcal{Q}}_0,\mathcal{R} \} , \mathcal{Q}_0+\hat{\mathcal{Q}}_0] =- [ \{ \mathcal{Q}_0,\mathcal{R} \} , \mathcal{Q}_0+\hat{\mathcal{Q}}_0] \nonumber \\ 
&=&-i[ S, \mathcal{Q}_0+\mathcal{Q}^0] =0 \,.
\eea{mass22}
For a generic operator $\mathcal{O}$ the condition for the existence of the limit is, likewise,
 $[\{ \mathcal{Q}^0,\mathcal{R} \} , O]=0$.  
 
\section{Intertwiner proof of the bosonic string no-ghost theorem}\label{app:B}

There exist many proofs of the no ghost theorem in bosonic string theory. A review of the original proof due to Brower~\cite{Brower:1972wj}, which uses the DDF operators~\cite{DelGiudice:1971yjh}, can be found in section 2.3 of~\cite{Green:2012gsw}. 
A different proof that relies on cohomological methods and properties of the BRST charge is described in section 4.4
of~\cite{Polchinski:1998rq}. \\

In this Appendix, we will revisit Polchinski's proof and use the intertwiner construction described in~\cite{Grassi:2024vkb} to prove the 
isomorphism of two cohomologies: a) the cohomology of the exact BRST operator of the bosonic string, b) the cohomology 
of a simplified BRST operator. The latter can be computed explicitly using the methods of~\cite{Green:2012gsw} and a few 
additional considerations.
The intertwiner we use was constructed explicitly 
in~\cite{Aisaka:2004xs} and by solving an evolution equation in~\cite{Erler:2021lzh}. It was used
by~\cite{Furuuchi:2006zq} to prove the no-ghost theorem of the bosonic string. This
appendix re-derives the proof of~\cite{Furuuchi:2006zq}, which is not widely known,
and emphasizes a key point central to our construction, namely that the
{\em existence} of the intertwiner can be proven without the need to know
explictly its form. While finding a formula for the intertwiner in terms of
oscillators is not difficult in the
case of the bosonic string~\cite{Aisaka:2004xs,Furuuchi:2006zq,Erler:2021lzh}, the intertwiner is
prohibitively
complex in more sophisticated theories, such as perturbative quantum gravity in
four dimensions. \\

The existence of an intertwiner allows for the construction of the physical Hilbert space of the bosonic string in terms of 
new dressed transverse operator that play the role of the DDF operators.
The proof  reviewed here also corrects a (nonfatal) 
mistake in Polchinski's proof.

\subsection*{The BRST charge of the bosonic string}
The coordinates of the bosonic string are two light-cone bosons, $X^+, X^-$, and transverse bosons $X^i$ $i=2,..25$. The transverse coordinates can be replaced by a
unitary conformal field theory with central charge $c=24$. 
The mode expansion {and commutation relations relevant} for our computation are
\beq
X^\pm(z)= x^\pm -{i\over 2} \alpha_0^\pm \log z + {i\over 2} \sum_{n\neq 0} {1\over n} \alpha_n^\pm z^{-n} ,\quad
[\alpha_m^+,\alpha_n^-]= -m\delta_{m+n,0} .
\eeq{1}
We'll confine our study to the holomorphic sector of the closed string, to avoid useless repetitions. The full holomorphic BRST charge of the bosonic string is written in terms
of the matter fields and a pair of ghosts, $c,b$. The scaling dimension of $c$ is $-1$ 
while the dimension of $b$ is $2$. Their
mode expansion and anticommutation relations are 
\beq
c(z)=c_n z^{-n+1}, \quad b(z)= b_n z^{-n-2} ,\quad \{c_n, b_m\}= \delta_{n+m,0} .
\eeq{2}
The ghost number operator $N_G$ assigns ghost number $+1$ to $c_n$ and $-1$ to $b_n$ for all $n$ so it obeys 
$[N_G , c_n] = +c_n$, $[ N_G , b_b]=-b_n$. The BRST charge $Q^B$ has ghost number $+1$ and the generators
of the Virasoro algebra of the matter+ghost system are related to it by $[Q^B,b_n]_+= L_n$. We won't need the 
detailed form of $Q^B$ but only a few important properties.
\begin{enumerate}
\item Define the charge~\cite{Polchinski:1998rq}
\beq
N_{LC} = \sum_{n=1}^\infty {1\over n} (\alpha_{-n}^+ \alpha_n^- - \alpha_{-n}^- \alpha_n^+ ).
\eeq{3}
It counts the number of nonzero-frequency $\alpha^-$ oscillators minus the {nonzero-frequency} $\alpha^+$ oscillators.
\item 
The BRST charge decomposes into term of definite $N_{LC}$ charge as~\cite{Polchinski:1998rq}
\beq
Q^B= Q_1 +Q_0 +Q_{-1}, \quad [N_{LC}, Q_j]=j Q_j .
\eeq{4}
The term $Q_1$ is 
\beq
Q_1= -k^+ \sum_{n\neq 0} \alpha_{-n}^- c_n ; \quad \mbox{it obeys } Q_1^2=0 .
\eeq{5}
It is implicitly assumed that the light cone momentum $k^+$ is nonzero; this property can be always guaranteed by a wise
choice of light cone coordinates.
\item
Define a new 
anticommuting 
charge~\footnote{This is \underline{not} the operator $R$ introduced by Polchinski in section 4.4 of~\cite{Polchinski:1998rq};
it was instead introduced in~\cite{Aisaka:2004xs} and used in~\cite{Aisaka:2004xs,Furuuchi:2006zq,Erler:2021lzh}.}
\beq
R= -{1\over k^+} \sum_{n\neq 0} {1\over n} \alpha_{-n}^+ b_n .
\eeq{6}
\item The charge $R$ defines a BRST homology
\beq
\{Q_1,R\}= S\equiv N_{LC} -N'_G, \quad N'_G=\sum_{n\neq 0} c_{n} b_{-n}.
\eeq{7}
$N'_G$ is the nonzero-frequency ghost number.
\end{enumerate}

An elementary consequence of the equation $\{Q^B,b_n\}=L_n$ is the decomposition
\beq
Q^B= c_0L_0 + \mbox{terms containing no }c_0.
\eeq{8}
Now Eqs.~(\ref{4},\ref{8}) imply that $Q^B$ decomposes into eigenstates of $S$ according to
\beq
Q^B= Q^0 +  Q^{-1} + Q^{-2}, \quad Q^0= Q_1 + c_0 L_0 , \quad [S,Q^n]=nQ^n . 
\eeq{9}
Notice that a) $S$ can be defined as $S=[Q^0,R]$, since $c_0L_0$ anticommutes with $R$ and b) $Q^0$ is
nilpotent, $(Q^0)^2=0$.

\subsection*{The intertwiner for the bosonic string}\label{appint1}
We define
\beq
Q(t)= Q^0 + Q_I(t), \quad Q_I(t)= \sum_{n=-1}^{-2} e^{nt} Q^n.
\eeq{mass18app}
The nilpotency of the BRST charge, $Q_B^2=0$, and the commutation relation $[S,Q^n]=nQ^n$ imply 
$\sum_{n+m=k} \{Q^n,Q^m\}=0$ for all 
$0\geq k \geq -4$ hence $Q(t)^2=0$ for all $t$ hence $Q^0 Q_I(t) + Q_I(t)Q^0 + Q_I(t)^2=0$.
Using the last identity we find
\bea
[\{Q_I(t),R\} ,Q(t)] &=& Q_I(t) R (Q^0+Q_I(t)) + R Q_I(t) (Q^0+Q_I(t))  \nonumber \\ && -(Q^0+Q_I(t)) Q_I(t)R   -(Q^0+Q_I(t))RQ_I(t) \nonumber \\ &=&
Q_I(t)RQ^0 +Q_I(t)RQ_I(t) -RQ^0Q_I(t) +Q_I(t)Q^0R  \nonumber \\ && -Q^0RQ_I(t) -Q_I(t)RQ_I(t) \nonumber \\ &=&
[Q_I(t), \{Q^0,R\}]= [Q_I(t),S]=- \sum_{0> n\geq -4}e^{nt} n Q_n= -{dQ_I\over dt} .
\eea{mass14app}
We define next an evolution operator $\Omega(t)$ by
\beq
{d\over dt} \Omega(t)= -  \{Q_I(t),R\} \Omega(t)
\eeq{}
with initial condition $\Omega(0)=1$. Using $\Omega(t)$ the charge $Q(t)$ can be written as 
\beq
Q(t)= \Omega(t) Q(0) \Omega^{-1}(t) = \Omega(t) Q(0) \Omega^\dagger(t) , \quad Q(0)=Q^B .
\eeq{mass20app}
The operator $\Omega(t)$ is unitary~\footnote{Since we did not define yet a scalar product, by ``unitary'' ``Hermitian'' etc. 
we mean here operators satisfying $U^{-1}= U^\dagger$, $H^\dagger = H$ etc. The adjoint $\dagger$ is an antilinear map acting on products of operators 
as $(AB)^\dagger= B^\dagger A^\dagger$ and defined on all matter, ghost and antighost modes, $x_n$, as
$x_{n}^\dagger = x_{-n}$.}
 and has a well defined $t\rightarrow +\infty$ limit because $\{Q_I(t),R\}$ is anti-Hermitian and decays exponentially in $t$. 
In fact we can say more. The operators $Q^0$, $R$ and $Q_I(t)$ commute with the level operator 
\beq
N= \sum_{n=1}^\infty \eta_{\mu\nu} \alpha_{-n}^\mu \alpha_n^\nu + n b_{-n} c_n + n c_{-n} b_{n} ,
\eeq{level}
 so $\Omega(t)$ is block-diagonal on the eigenspaces of $N$, 
 where it is a finite-dimensional matrix. \\
 
The evolution operator $\exp(tS)$ instead has an ill-defined limit $t\rightarrow +\infty$ on the 
subspace spanned by state $\Psi$ of negative $S$-charge, $S\Psi=(N_{LC}- N'_G)\Psi=s\Psi$, $s<0$. In fact 
even the $t\rightarrow +\infty$ limit of the conjugation 
$\exp(-tS) O \exp(tS)$ is ill-defined on operators of negative $S$ charge. \\

At $t=+\infty$ $Q(+\infty)=Q^0$ so $\Omega(+\infty)\equiv \Omega$ is an intertwiner between the exact BRST charge
$Q^B$ and the simple charge $Q^0$
\beq
Q^0 \Omega = \Omega Q^B .
\eeq{mass21app}
The construction of $\Omega$ requires that the BRST charge is nilpotent, which is true if and only if the central charge 
of the matter theory is $c=26$.

\subsection*{The $Q^0$ cohomology}
The existence of a unitary intertwiner guarantees that the BRST cohomology of $Q^B$, $H(Q^B)$ and
the cohomology of $Q^0$, $H(Q^0)$ are isomorphic. The $Q^0$ operator is the sum of two term $Q_1$, which acts
on the nonzero modes of $\alpha_n^\pm$, $b_n,c_n$, $n\neq 0$ and $c_0L_0$. $L_0$ commutes with $Q_1$,
$R$, $c_0$ and all the $Q^n$. So the evolution operator $\Omega(t)$ and the cohomologies leave the eigenspaces of
$L_0$ invariant. On an eigenspace $\mathcal{H}_X$ defined by $L_0v=Xv$ $\forall v\in \mathcal{H}_X$ the cohomology
is 
\beq
Q^0=Q_1+Xc_0 .
\eeq{10}
The Hilbert space on which $Q^0$  acts is a product $\mathcal{H}_F \otimes \mathcal{H}_0$. $\mathcal{H}_F$ is the Fock space of the nonzero oscillators $\alpha_n^\pm$, $b_n,c_n$, $n\neq 0$ while $\mathcal{H}_0$ is 
the Fock space of $b_0$, $c_0$. It is the linear span of the states $\Psi_{\downarrow}$, defined by 
$b_0\Psi_{\downarrow}=0$, and $\Psi_{\uparrow}=c_0 \Psi_{\downarrow}$. 
K\"unneth formula~\cite{Nitti:2003ef,Griffiths:1978hc} is a general result in cohomology stating that the sum of two cohomologies, $Q+Q'$, 
acting on the product Hilbert space $\mathcal{H} \otimes \mathcal{H}'$ is the tensor product of the two cohomologies
$H(Q)\otimes H(Q')$. In our case $H(c_0X)$ is empty except for $X=0$, where it's the whole
$\mathcal{H}_0$. So $H(Q^0)$ is either empty, for $X\neq 0$ or equal to $H(Q_1)\otimes \mathcal{H}_0$, for $X=0$. 

In fact the result  for $X=0$ is tautological and the calculation of the cohomology for $X\neq 0$ is elementary 
because any state in the cohomology must be $Q_1$-closed and all such states can be written as linear combinations of the product states 
 $V \otimes \Psi_{\downarrow}$ and $W \otimes \Psi_{\uparrow}$, with $Q_1V=Q_1W=0$. On the other hand,
 \beq
 W \otimes \Psi_{\uparrow} = {1\over X} Q^0 (W \otimes b_0 \Psi_{\uparrow}), \qquad
 Q^0 (V \otimes  \Psi_{\downarrow}) = X V \otimes \Psi_{\uparrow}.
 \eeq{11}
 The first state is exact and the second is not closed; neither belong to the cohomology, which is therefore empty. \\
 
 Next we must define a scalar product. We can define it to be nonzero either on  the subspace 
 $\mathcal{H}' \otimes \Psi_{\uparrow}$ or on $\mathcal{H}' \otimes \Psi_{\downarrow}$. We denoted with $\mathcal{H}'$
 the space of all string modes except $b_0,c_0$, but including the spacetime momenta $k^\mu$. 
 To start with, we define ground states of the nonzero oscillators {and of $\alpha_0^\mu$} by
  \beq
   \alpha_n^\mu \Phi_k=0 , \; n\geq 1, \quad \alpha_0^\mu \Phi_k  = {1\over 2}k^\mu \Phi_k , \quad 
  \quad b_n \Phi_k  =0 \; n\geq 1, \quad c_n \Phi_k =0 , \; n\geq 1 .
  \eeq{14}
 Then $\mathcal{H}'$ is a Fock space built in the usual manner from the ground states $\Phi_k$. An indefinite-metric
 scalar product is then defined by the properties
 \bea
 <\alpha_{-n} V,W> &=& <V,\alpha_n W>, \quad <c_{-n} V,W>=<V,c_nW>, \quad <b_{-n} V,W>=<V,b_nW>, \nonumber \\
 <\Phi_k , \Phi_q> &=& (2\pi)^{26} \delta^{26}(k-q) , \quad \forall \; V,W\in \mathcal{H}' .
 \eea{14a}
  The  scalar product on $\mathcal{H}_0$ is defined on the basis vectors as 
  \beq
  \langle \Psi_\downarrow ,\Psi_\uparrow\rangle = \langle \Psi_\uparrow ,\Psi_\downarrow \rangle =1, \quad 
  \langle \Psi_\uparrow ,\Psi_\uparrow \rangle = \langle \Psi_\downarrow ,\Psi_\downarrow \rangle=0 .
  \eeq{14b}
  We define the scalar product on $\mathcal{H}' \otimes \mathcal{H}_0$  using the bilinear form
  \beq
  (V\otimes v , W\otimes w)=
  <V, W> \langle v,w \rangle , \quad V,W\in\mathcal{H}' , \quad v,w\otimes \mathcal{H}_0 .
  \eeq{14c}
  Finally, the two choices for the scalar product are 
  \beq
 a)\;\; (V, b_0 \delta(L_0)  W), \quad b)\;\; (V,  c_0 W), \quad V,W\in \mathcal{H}' \otimes \mathcal{H}_0.
 \eeq{12}
 The spectrum of the the operator $L_0$ is continuous because of the momentum contribution 
 $L_0= {1\over 8} k^2 + .... $. Only choice $a)$ defines a Hilbert space of normalizable
 states on the cohomology $H(Q_1)$, the space of {\em coinvariants}~\footnote{This construction has been used 
 in~\cite{Chandrasekaran:2022cip,Marolf:2008hg} for constructing physical states in quantum gravity in de Sitter space.}.   
 
 In string theory the standard choice is instead to project over the subspace $\mathcal{H}' \otimes \Psi_{\downarrow}$ 
 by selecting the scalar product $(V,  c_0 W)$. With this choice the BRST cohomology is still $H(Q_1)$ but its vectors are 
 only plane-wave normalizable. 
 
 \subsection*{The physical Hilbert space}
  A basis of representatives of the cohomology $H(Q^0)$ is
  \beq
  \Psi_{phys}\equiv \prod_I \alpha^{i_I}_{n_I} \Phi_k \otimes \Psi_{\downarrow} , \;   2\leq i_I \leq 25\; \forall i_I. 
  \eeq{13}
  The momenta $k^\mu$ are not independent of the transverse-oscillator level number 
  \beq
  N^T=\sum_{i=2}^{25} \sum_{n>0} \alpha^i_n \alpha^i_{-n}. 
  \eeq{13a}
  They are linked by the condition 
  \beq
  L_0\Psi_{phys} =0 \Rightarrow  \left( {1\over 8} k^2 + N^T -1 \right) \Psi_{phys} =0 .
  \eeq{15}
  {\em Notice that this condition is not part of the definition of the cohomology $H(Q_1)$}, which does not involve 
  any transverse momenta. Ref.~\cite{Polchinski:1998rq} claims to show that $H(Q_1)$ is isomorphic to $H(Q^B)$. 
  This is clearly impossible unless
  Eq.~\eqref{15} holds. In other words, the correct statement is that  $H(Q^0)$  is isomorphic to 
  $H(Q^B)$. It is the condition $b_0 \Psi=0$ --or, equivalently, the choice of scalar product  $b)$ in~\eqref{12}-- that
  reduces $H(Q^0)$ to $H(Q_1)$ {\em on the subspace}  $L_0\Psi_{phys}=0$. \\

  The positivity of the scalar product is evident for the states~\eqref{13}. The intertwiner $\Omega$ is  unitary 
  in the metric $(V,  c_0 W)$ and $\Omega^\dagger V \in H(Q^B) \iff V\in H(Q^0)$, so
  \beq
  (\Omega^\dagger  V,  c_0 \Omega^\dagger W) = ( V,  \Omega c_0 \Omega^\dagger  W) = (V,  c_0 W) >0\quad
  \forall V,W \in H(Q^0).
  \eeq{16}
  
 The operators $Q_I(t)$ and $R$ are odd under fermionic parity and commute with the level operator $N$ 
 given in eq.~\eqref{level}.
  The ground states 
 $\Phi_k\otimes \Psi_\downarrow$  obey $N\Phi_k\otimes \Psi_\downarrow=0$. Because of fermionic parity, both
 $Q_I(t) \Phi_k\otimes  \Psi_{\downarrow}$ and $R \Phi_k \otimes \Psi_{\downarrow}$ must be proportional to the
 only other states annihilated by $N$, namely $\Phi_k \otimes c_0 \Psi_\downarrow$. On the other hand, neither
 $Q_I(t)$ nor $R$ contain $c_0$, so they annihilate $\Phi_k\otimes \Psi_\downarrow$. The same is therefore true of the
 evolution operator $[Q_I(t),R]_+$ and of $\Omega(t)$.  
  We can then generate the physical Hilbert space using dressed transverse-boson creation operators. 
 Specifically, we generate states in the $Q^0$ cohomology by applying  to the ground state 
 $\Phi_k\otimes \Psi_\downarrow$ the operators 
 \beq
 a^i_{-n} e^{2in \alpha_0^+}, \quad i=2,...,25, \quad n>0.
 \eeq{17}
 To satisfy eq.~\eqref{15} we choose $k_+=2,$ $k_-=-2,$ $k^i=0$.
 The operators~\eqref{17} play the same role as the DDF vertex operators in light-cone quantization~\cite{DelGiudice:1971yjh}. 
 The $Q^B$ cohomology is then generated by the
 dressed DDF-like operators  
 \beq
 A^i = \Omega^\dagger a^i_{-n} e^{2in \alpha_0^+} \Omega .
 \eeq{18}

\newpage
\bibliographystyle{nb}
\bibliography{references}
\end{document}